\PassOptionsToPackage{hypertexnames=false}{hyperref}
\documentclass[pdflatex,bst/sn-mathphys-num]{sn-jnl}
\usepackage[utf8]{inputenc}%
\usepackage[T1]{fontenc}%
\usepackage{graphicx}%
\usepackage{multirow}%
\usepackage{amsmath,amssymb,amsfonts}%
\usepackage{bm}%
\usepackage{amsthm}%
\usepackage{mathrsfs}%
\usepackage[title]{appendix}%
\usepackage{xcolor}%
\usepackage{textcomp}%
\usepackage{manyfoot}%
\usepackage{booktabs}%
\usepackage{algorithm}%
\usepackage{algorithmicx}%
\usepackage{algpseudocode}%
\usepackage{listings}%
\usepackage{float}
\usepackage{booktabs}       
\usepackage[table]{xcolor} 
\usepackage{siunitx}        
\usepackage{array}         
\usepackage{colortbl}

\theoremstyle{thmstyleone}%
\theoremstyle{thmstyletwo}%
\theoremstyle{thmstylethree}%
\begin{document}

\title{Inverse-designed meta processing units for multi-task near-field photonic computing}

\author[1]{\fnm{Chu} \sur{Wu}}\email{wuc23@mails.tsinghua.edu.cn}
\author[1]{\fnm{Zeyu} \sur{Cai}}\email{caizy25@mails.tsinghua.edu.cn}
\author[1]{\fnm{Songtao} \sur{Yang}}\email{yang-st24@mails.tsinghua.edu.cn}
\author[2]{\fnm{Ruoyu} \sur{Shen}}\email{shenruoyu@zjlab.ac.cn}
\author[1]{\fnm{Yinan}\sur{Zhao}}\email{zhaoyinan@mail.tsinghua.edu.cn}
\author[1]{\fnm{Haiou} \sur{Zhang}}\email{azhho@mail.tsinghua.edu.cn}
\author[2]{\fnm{Wei} \sur{Chu}}\email{chuwei@zjlab.ac.cn}
\author*[1]{\fnm{Xing} \sur{Lin}}\email{lin-x@tsinghua.edu.cn}

\affil*[1]{\orgdiv{Department of Electronic Engineering}, \orgname{Tsinghua University}, \orgaddress{\street{30 Shuangqing Road}, \city{Beijing}, \postcode{100084}, \country{China}}}
\affil[2]{\orgdiv{Key Laboratory of Photonic Integration and Quantum Information}, \orgname{Zhangjiang Laboratory}, \orgaddress{\street{100 Haike Road}, \city{Shanghai}, \postcode{201210}, \country{China}}}

\abstract{
Integrated photonic neural networks require optical operators that are simultaneously compact,
matrix-general and compatible with task-level reconfigurability. Here we introduce a
meta processing unit (MPU), an inverse-designed near-field photonic device that implements
local complex matrix transformations within a shallow-etched silicon region. Each
2$\times$2 operator occupies 9.6$\,\mu\mathrm{m}\times$4.8$\,\mu\mathrm{m}$ and is designed as a
reusable passive matrix primitive that can be combined with reconfigurable MZI neurons.
We demonstrate a 3-bit quantized MZI-equivalent unitary device library with an effective
reconstruction precision of 3.32 bits.
Beyond unitary operators, we validate arbitrary complex 2$\times$2 matrix fitting and a
cascaded 4$\times$4 matrix operation with 92.7\% fidelity. We further integrate the MPU
with active photonic components and hardware-in-the-loop training, achieving test
accuracies of 83.5\% and 80.9\% on dual-task vowel recognition. In large-scale EMNIST
simulations, a fine-grained neuron-level MPU replacement strategy reaches 87.64\%
average accuracy at 90\% shared-MPU replacement, outperforming a layer-level baseline by
7.26 percentage points. These results establish inverse-designed MPUs as compact passive
matrix operators for heterogeneous, hardware-adaptive photonic neural networks.
}


\maketitle

%%%%% Introduction %%%%%%%%%%%%%%%%%%%%%%%%%%%%%%%%%%%%%%%%%%%%%%%%%
\section{Introduction}\label{sec:intro}
\begin{figure}[H]
\centering
\includegraphics[width=0.9\textwidth]{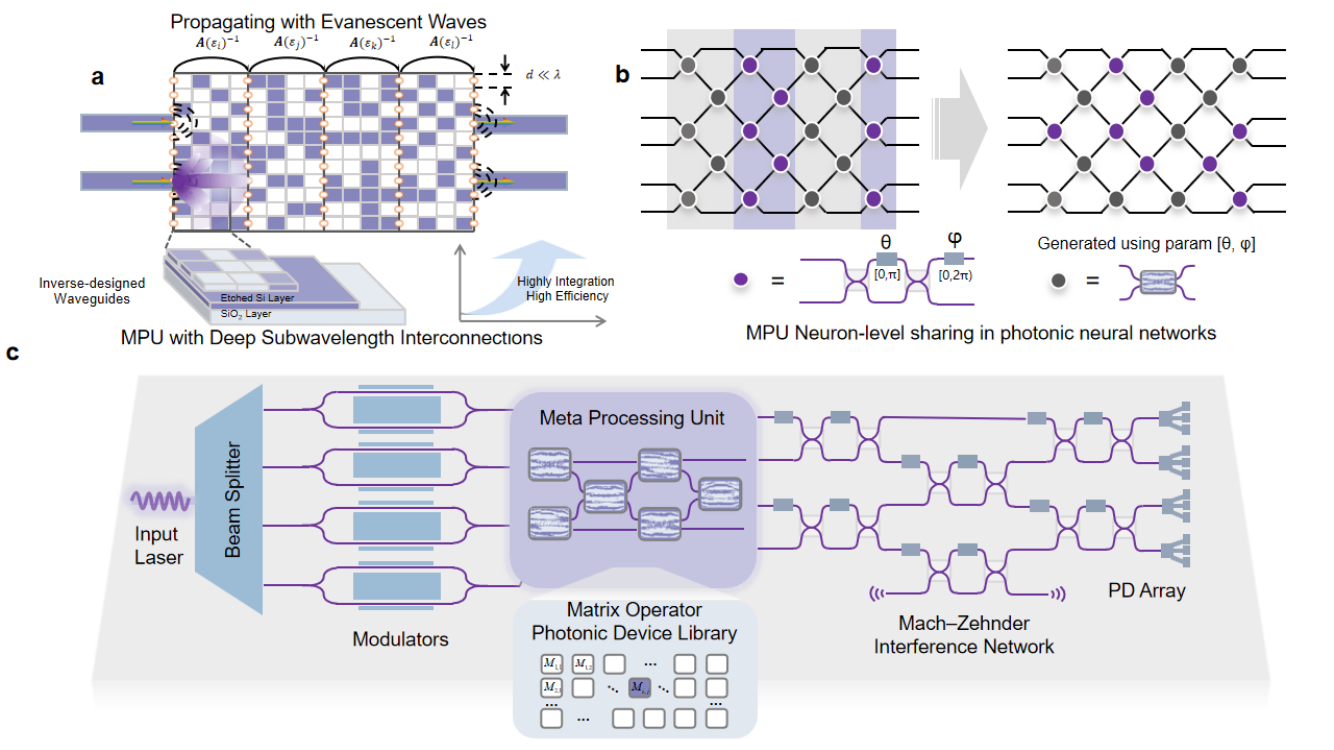}
\caption{\textbf{Inverse-designed MPUs with heterogeneous device integration for high-density multi-task photonic neural network.} (a) Device schematic of the MPU with deep subwavelength interconnections, showing the inverse-designed waveguide array, evanescent wave coupling mechanism. (b) Schematic of the task-aware neuron-level MPU sharing scheme for photonic neural networks (PNNs), where active reconfigurable MZI neurons (purple) and task-shared inverse-designed MPU neurons (gray) are dynamically allocated to optimize hardware efficiency for multi-task inference. (c) Schematic of the full integrated heterogeneous photonic computing system, consisting of an input laser, beam splitters, modulator array, MPUs with a pre-designed matrix operator photonic device library, cascaded Mach–Zehnder interferometer (MZI) network, and photodetector (PD) array.}
\label{fig1}
\end{figure}

Photonic computing exploits the fundamental parallelism and low-latency propagation of electromagnetic waves to accelerate artificial intelligence (AI) workloads\cite{shastri2021photonics_ai,markovic2020physics_neuromorphic,wetzstein2020deep_optics}. Recent integrated photonic processors have further demonstrated rapid progress in nanophotonic media, on-chip photonic neural networks and all-optical intelligent processing\cite{zhao2025nanophotonic_media,gu2025all_integrated_sensing,wu2025scaling_on_chip_pnn,chen2025all_optical_synthesis}. However, translating foundational optical physics into scalable machine learning hardware remains severely bottlenecked by the inherent trade-off between sub-wavelength integration density and computational reconfigurability\cite{bogaerts2020programmable_photonic_circuits,nikkhah2024inverse_designed_vmm,du2024ultracompact_multifunctional,liu2025ultracompact_multitask}.

Early programmable photonic accelerators relied on classical cascaded Mach--Zehnder interferometer (MZI) meshes\cite{bogaerts2020programmable_photonic_circuits,shen2017deep_learning_coherent,clements2016optimal_multiport,reck1994universal_unitary,harris2018linear_programmable,carolan2015universal_linear_optics}. While these provide theoretically rigorous unitary operations, their extensive $\mathcal{O}(N^2)$ active footprint and compounding thermal crosstalk fundamentally restrict scalable edge deployment\cite{bogaerts2020programmable_photonic_circuits,clements2016optimal_multiport,reck1994universal_unitary,carolan2015universal_linear_optics,bandyopadhyay2021hardware_error_correction}. Conversely, passive on-chip diffractive optical networks achieve high miniaturization but lack dynamic reconfigurability for multi-task execution\cite{lin2018diffractive_dnn,zhou2021reconfigurable_dpu,zhu2022integrated_chip_dnn,fu2023on_chip_diffractive,liu2025ultracompact_multitask}. Recent large-scale heterogeneous photonic chiplets, e.g., Taichi\cite{xu2024taichi}, have combined passive diffractive modules with active interferometric structures to improve system-scale photonic computing. However, in such architectures, passive diffractive components are typically deployed as fixed layer- or block-level operators, while reconfigurability is mainly introduced at a relatively coarse architectural granularity. This limits their ability to selectively allocate task-shared and task-aware optical transformations at the level of individual computing neurons. At the microscopic device level, standard inverse design has enabled highly miniaturized operators\cite{nikkhah2024inverse_designed_vmm,sun2025edge_guided_inverse_design,molesky2018inverse_design,minkov2020automatic_differentiation,piggott2015wavelength_demux}. However, many inverse-designed photonic components are optimized as isolated application-specific devices, and their direct use as reusable, coherently cascaded matrix operators remains non-trivial.

Here, we introduce an inverse-designed meta processing unit (MPU) as a compact passive optical operator for constructing high-density photonic neural networks. By using a shallow-etched near-field diffractive structure, the MPU maintains predominantly forward optical propagation and implements deterministic $2\times2$ transformations within an ultra-compact $9.6\,\mu\text{m} \times 4.8\,\mu\text{m}$ footprint. Rather than treating inverse-designed components as isolated application-specific devices, we formulate MPUs as reusable optical operator units that can be cascaded and integrated with reconfigurable MZI neurons. This MPU-MZI heterogeneous architecture enables task-shared passive operators and task-private reconfigurable neurons to be allocated at the level of individual optical computing units, providing a compact route toward scalable multi-task photonic neural networks. The resulting operator library samples the MZI-equivalent unitary space on a 3-bit
\(\theta\)-\(\phi\) grid.

Building on this MPU-MZI heterogeneous architecture, we further explore task-aware neuron-level MPU replacement for multi-task photonic learning\cite{ruder2017multitask_overview}. By identifying optical neurons whose transformations can be shared across related tasks, selected reconfigurable MZI units are consolidated into task-shared passive MPUs, while a sparse set of task-private MZI neurons is retained to provide task-specific adaptation. This strategy is designed to reduce footprint and control overhead while preserving the optical degrees of freedom most relevant to multi-task inference. We validate the proposed framework through on-chip dual-task vowel recognition on an active 130-nm silicon-on-insulator (SOI) processor and large-scale EMNIST simulations for digit, uppercase-letter, and lowercase-letter classification\cite{davis1980vowel_features,cohen2017emnist}. These results show that MPU-MZI heterogeneous integration can support both experimentally measured multi-task photonic inference and scalable task-aware operator sharing in larger photonic neural networks.

%%%%% Results %%%%%%%%%%%%%%%%%%%%%%%%%%%%%%%%%%%%%%%%%%%%%%%
\section{Results}

\subsection{Inverse-designed MPU as a compact complex optical matrix operator}

The basic physical unit of the proposed architecture is an inverse-designed meta processing
unit (MPU), which implements a compact complex-valued optical matrix operation within a
sub-wavelength patterned silicon region. In contrast to conventional programmable
Mach-Zehnder interferometer (MZI) meshes, where a target matrix is decomposed into a
large sequence of tunable interferometers, the MPU directly maps the desired input-output
scattering relation onto a passive nanophotonic structure. The computational variable is
therefore the complex modal amplitude carried by each single-mode waveguide. For a
two-input-two-output operator, the incident and outgoing fields are written as
\begin{equation}
    \mathbf{y}
    =
    \mathbf{M}\mathbf{x},
    \qquad
    \mathbf{M}
    =
    \begin{bmatrix}
    a_{11} & a_{12}\\
    a_{21} & a_{22}
    \end{bmatrix},
\end{equation}
where $a_{ij}$ denotes the complex transmission coefficient from input port $j$ to output
port $i$. This field-level formulation is essential for coherent photonic computing, because
the output of one MPU can be directly cascaded into subsequent optical modules without
being reduced to an intensity-only response.

\begin{figure}[ht]
\centering
\includegraphics[width=1.0\textwidth]{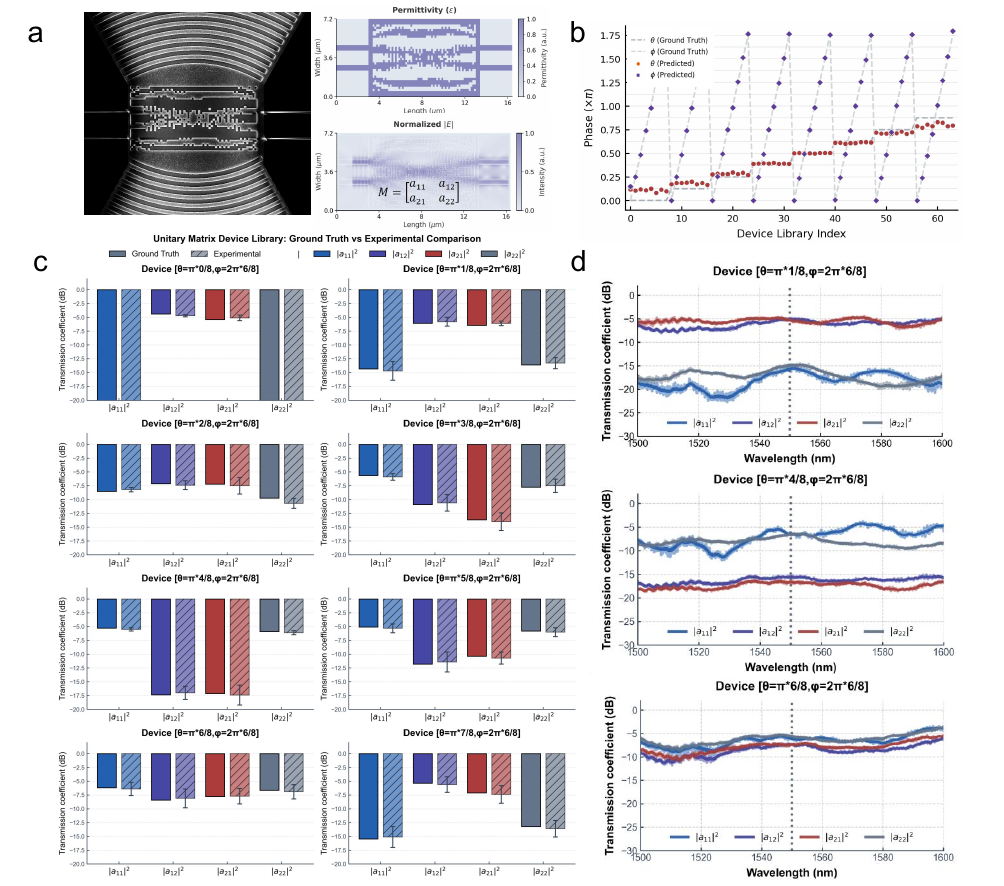}
\caption{
\textbf{Statistical performance of the inverse-designed $2\times2$ unitary MPU library.}
\textbf{a,} SEM image of a fabricated shallow-etched $2\times2$ MPU and the corresponding
inverse-designed topology and measured optical intensity distribution at 1550 nm.
\textbf{b,} target and experimentally reconstructed \(\theta-\phi\) states
\textbf{c,} Normalized transmission coefficients of four matrix elements for eight
representative inverse-designed unitary operators. Error bars denote the standard deviation
of repeated measurements (\(N=5\)).
\textbf{d,} Wavelength-dependent transmission coefficients of three representative devices
across the 1500-1600 nm band.
}
\label{fig2}
\end{figure}

Figure~\ref{fig2}a shows a representative fabricated MPU and its corresponding
inverse-designed topology. Each operator occupies only
$9.6~\mu\mathrm{m}\times4.8~\mu\mathrm{m}$ and is formed by a shallow-etched silicon
nanostructure with a binary pixelated design grid. The internal pattern is parameterized by
$160~\mathrm{nm}$ square features, providing a lithography-compatible design space for
controlling the local effective permittivity. A shallow etch depth of $70~\mathrm{nm}$ is used
to reduce back-scattering while preserving sufficient wavefront modulation strength. This
geometry produces predominantly forward optical propagation and enables the diffractive
region to act as a compact, reusable optical matrix primitive.

The measured near-field intensity distribution in Fig.~\ref{fig2}a confirms that the designed
wavefront is confined within the MPU and couples to the intended output waveguide modes.
The absorbing or weakly scattering boundary structures suppress residual uncollected fields
and reduce parasitic feedback into the computational ports. In simulations, representative
identity-like benchmark operators can reach transmission efficiencies approaching
$80\%$ when the design region provides sufficient spatial degrees of freedom
(Supplementary Fig.~S2). This transmission level is markedly higher than typical
ultra-compact metaline-type diffractive operators, where single-device transmission is often
only at the percent or sub-percent level. The MPU therefore provides a practical compromise
between compact footprint and usable optical throughput, which is a prerequisite for
cascaded photonic computation.

This design principle converts inverse-designed nanophotonics from isolated
application-specific components into a reusable matrix-operator library. Passive MPUs
supply dense fixed complex transformations, while a smaller number of active MZI or
phase-tuning elements provide calibration and task-dependent adaptability. This forms the
device-level foundation of the heterogeneous MPU-MZI photonic computing architecture.

\subsection{Quantized unitary MPU library and experimental validation}

We first validate the MPU concept using a digitally indexed unitary matrix library. The
ground-truth operators are generated from the standard MZI-equivalent parameterization,
\begin{equation}
    \mathbf{U}(\theta,\phi)
    =
    i e^{i\theta/2}
    \begin{bmatrix}
    e^{i\phi}\sin(\theta/2) & e^{i\phi}\cos(\theta/2)\\
    \cos(\theta/2) & -\sin(\theta/2)
    \end{bmatrix},
    \label{eq:mzi_unitary_result}
\end{equation}
where $\theta$ controls the splitting ratio and $\phi$ sets the relative phase. Both
parameters are quantized using a 3-bit grid, generating an $8\times8$ set of
MZI-equivalent unitary transformations. For each quantized \((\theta,\phi)\) pair, an
individual MPU is inverse-designed to reproduce the corresponding complex scattering
matrix.

Figure~\ref{fig2}b compares the target and experimentally extracted \(\theta-\phi\) parameter
pairs across the device library. The measured states follow the designed quantized grid,
showing that the passive MPU can faithfully reproduce the intended MZI-equivalent
operator states after fabrication. This agreement is further supported by the amplitude
statistics in Fig.~\ref{fig2}c, where representative devices show consistent normalized
transmission coefficients across the four matrix elements. The measured insertion losses
of the passive unitary operators are in the range of \(4.8\)-\(5.4~\mathrm{dB}\), which is
substantially lower than the effective loss expected from many ultra-compact diffractive
metaline operators with percent-level transmission. Importantly, the measured responses
retain the designed relative amplitude contrast among the four ports, rather than merely
maximizing total transmitted power.

To quantify the analog reconstruction quality of the unitary operator library, we convert the
normalized complex-matrix error into an effective bit precision. Across the experimentally
measured MPU library, the reconstructed operators show an effective precision of
\(3.32\) bits. This value should be distinguished from the 3-bit digital grid used to index the
\(\theta-\phi\) targets: the quantization grid defines the intended operator states, whereas
the effective bit precision evaluates how accurately the fabricated nanophotonic devices
reproduce their assigned complex matrices. The result indicates that the inverse-designed
MPUs do not simply approximate a coarse switching table, but preserve analog
complex-valued matrix information with useful precision.

Figure~\ref{fig2}d further evaluates the spectral behavior of representative devices across
the 1500-1600 nm communication band. The measured transmission coefficients preserve
their designed contrast around the 1550 nm operating wavelength and show stable
device-dependent spectral trends. This wavelength robustness is important for practical
operation because it provides tolerance against laser drift, fabrication-induced spectral shift
and moderate wavelength-dependent coupling variations. Together, Fig.~\ref{fig2}
establishes that the MPU can serve as a compact passive library of experimentally
reproducible unitary optical matrix operators.

\begin{figure}[ht!]
\centering
\includegraphics[width=1.0\textwidth]{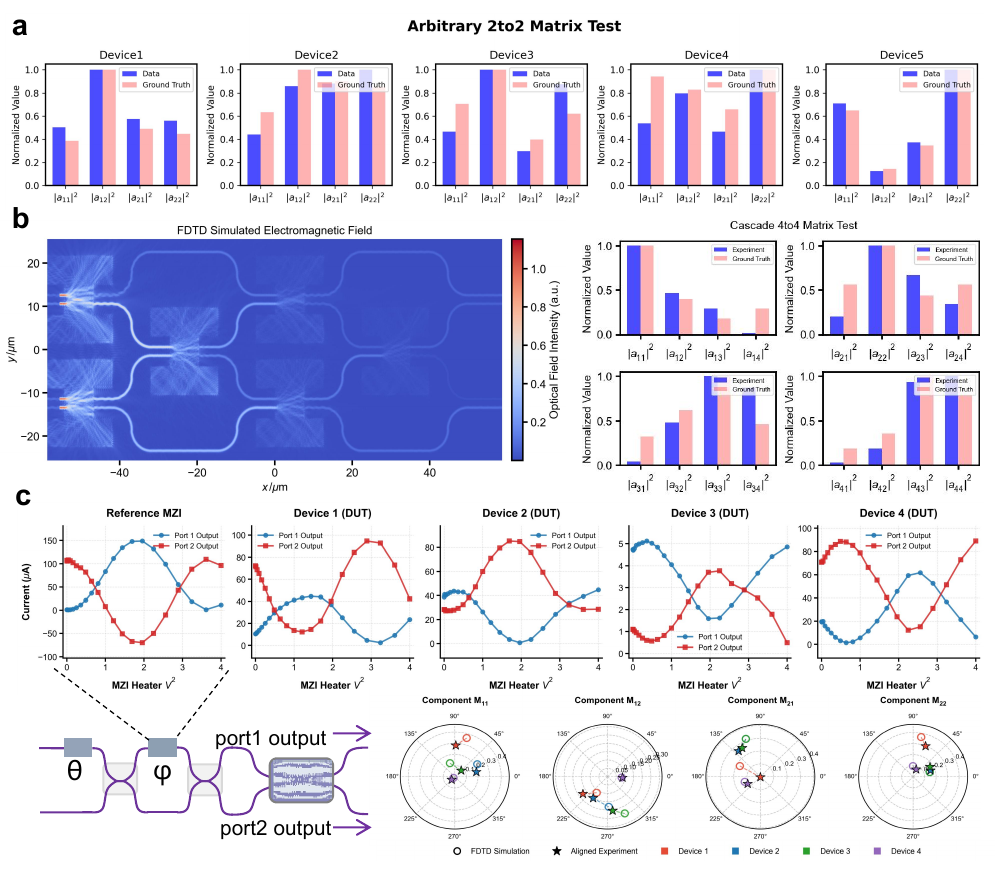}
\caption{
\textbf{MPW validation of arbitrary and cascaded complex MPU operators.}
\textbf{a,} Normalized transmission amplitudes of five fabricated arbitrary
$2\times2$ inverse-designed matrix operators, compared with their ground-truth targets.
\textbf{b,} Left: FDTD-simulated optical field distribution of a cascaded $4\times4$ dense
matrix operator assembled from local MPU blocks. Right: Normalized transmission
amplitudes of the reconstructed $4\times4$ transmission matrix.
\textbf{c,} Complex-domain calibration using on-chip MZI phase scanning. Top and
bottom-left: Measured photocurrent responses of a reference MZI and four cascaded
MZI-MPU devices under voltage scanning. Bottom-right: Polar plots comparing the
FDTD-simulated and experimentally reconstructed complex transmission coefficients.
}
\label{fig3}
\end{figure}

\subsection{Non-unitary matrix synthesis and cascaded complex-operator validation}

The quantized unitary library verifies that an MPU can reproduce MZI-equivalent
$2\times2$ transformations, but practical optical neural networks require a broader class
of operators. Neural-network weight matrices are generally non-unitary: their singular
values are not constrained to unity, and both amplitude and phase degrees of freedom are
needed. We therefore extended the experimental validation from unitary operators to
arbitrary complex-valued $2\times2$ matrices fabricated through the active silicon photonic
MPW process.

Figure~\ref{fig3}a shows five representative arbitrary $2\times2$ matrix operators. For
each device, the normalized amplitudes of the four transmission coefficients are compared
with the corresponding ground-truth targets. The measured responses reproduce the
relative amplitude distributions of different non-unitary matrices, including strongly
imbalanced matrix elements. This result confirms that the inverse-designed MPU is not
limited to energy-conserving MZI-like transformations, but can realize amplitude-weighted
complex matrix operators under a controlled loss budget. This capability is essential for
photonic neural-network accelerators, where the implemented operators must approximate
general learned weights rather than only unitary rotations.

We next tested whether local MPU operators can be composed into a larger dense optical
transformation. As shown in Fig.~\ref{fig3}b, multiple $2\times2$ MPU blocks were
assembled into a cascaded $4\times4$ matrix topology. The simulated optical field
distribution illustrates sequential field evolution through the compact diffractive operators
and waveguide interconnections. The reconstructed $4\times4$ transmission amplitudes
show good agreement with the designed dense matrix pattern, yielding an overall matrix
fidelity of $92.7\%$. The result indicates that local
inverse-designed operators can be cascaded to form higher-dimensional matrix
transformations while maintaining useful analog accuracy.

To verify that the MPU preserves the complex-valued operator structure rather than only
the output intensity, we further performed complex-domain calibration using on-chip MZI
phase scanning (Fig.~\ref{fig3}c). A reference MZI was first measured by sweeping its
phase-shifter voltage, producing a periodic photocurrent response at the output ports. When
an MPU is cascaded after the MZI, the output currents become
\begin{equation}
    I_k(V)
    =
    A_k
    +
    B_k
    \cos
    \left[
    \Delta\phi(V)+\varphi_k
    \right],
\end{equation}
where \(A_k\), \(B_k\) and \(\varphi_k\) are obtained from sinusoidal fitting, and
\(\Delta\phi(V)\) is the voltage-controlled phase of the reference MZI. By comparing the
cascaded MZI-MPU response with the reference MZI scan, the complex coefficient of each
output channel can be reconstructed as
\begin{equation}
    \widehat{M}_{kj}
    \propto
    \sqrt{\frac{B_k^{\mathrm{DUT}}}{B_k^{\mathrm{ref}}}}
    \exp
    \left[
    i
    \left(
    \varphi_k^{\mathrm{DUT}}
    -
    \varphi_k^{\mathrm{ref}}
    \right)
    \right],
\end{equation}
up to a global amplitude scale and a common phase offset. The detailed calibration and
alignment procedure is provided in the Supplementary Information Chapter S3.3.

The polar plots in Fig.~\ref{fig3}c compare the experimentally reconstructed complex
coefficients with the FDTD-simulated values for four fabricated devices. After global
alignment, the measured points follow the simulated amplitude-phase distributions,
demonstrating that the fabricated MPUs retain the relative complex matrix structure needed
for coherent cascading. These measurements show that the MPU is not merely a compact diffractive splitter, but a
coherently composable complex-valued matrix primitive that can be extended from local
$2\times2$ operators to larger dense optical transformations.
\begin{figure}[ht!]
\centering
\includegraphics[width=1.0\textwidth]{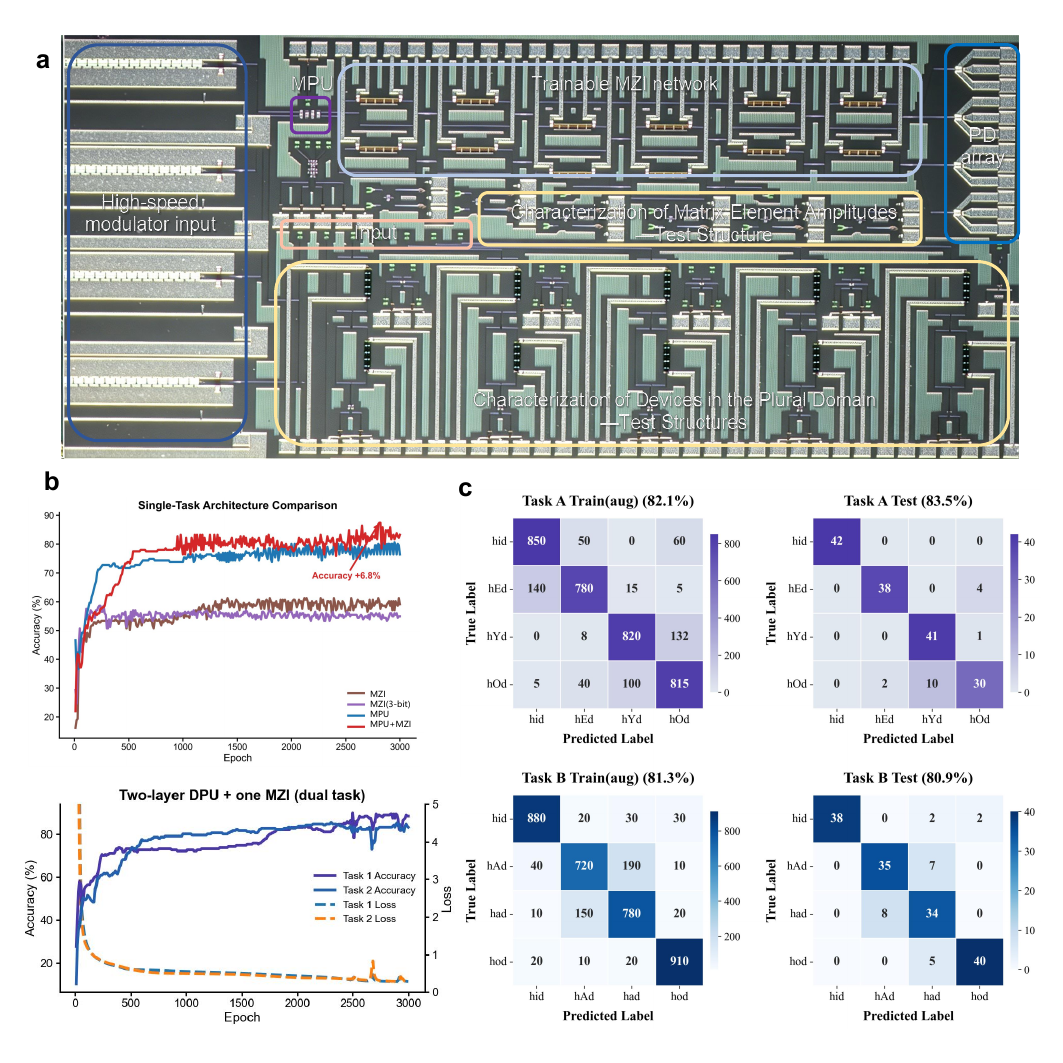}
\caption{\textbf{Experimental validation of the task-aware multi-task photonic computing system for vowel classification applications.} (a) Microscope image of the fully integrated photonic chip used for system-level testing of the multi-task vowel classification. (b) Top: Convergence curves of classification accuracy for single-task architectures. Bottom: Convergence curves of the dual-task learning workflow. (c) Confusion matrices of the training and test sets for two vowel classification tasks, achieving 83.5$\%$ and 80.9$\%$ test accuracy.}
\label{fig4}
\end{figure}

\subsection{Hardware-in-the-loop validation on dual-task vowel classification}

Having validated the MPU as a compact complex matrix operator, we next integrated it into
a task-level photonic neural network to evaluate whether the proposed heterogeneous
MPU-MZI architecture can support experimentally measured inference. Figure~\ref{fig4}a
shows the fabricated active photonic chip used for the hardware-in-the-loop experiment.
The chip integrates high-speed modulator inputs, an inverse-designed MPU block, a
trainable MZI network, and an on-chip photodetector array. This layout allows optical
features to be encoded by MZMs, processed by the hybrid passive-active photonic core,
and read out electrically through integrated photodetectors.

The experimental task is dual-task vowel classification. The input acoustic features are
first compressed by a lightweight digital PreNet and then mapped to calibrated MZM driving
voltages. The optical core performs the physical linear transformation, while the detected
photocurrents are normalized and passed to task-specific digital readout heads. The
trainable parameters include the digital PreNet weights, MZM input-scaling factors,
photonic phase settings, and detector readout calibration factors. Because the fabricated
chip contains nonidealities that are difficult to model exactly, including modulator
nonlinearity, optical-path imbalance, detector responsivity mismatch, coupling fluctuation
and thermal drift, we optimized the system directly with the measured hardware response.
The closed-loop update was performed using simultaneous perturbation stochastic
approximation (SPSA), which estimates a stochastic descent direction from two hardware
evaluations per iteration and is compatible with measured-response and in situ photonic training workflows\cite{spall1992multivariate_spsa,zheng2023dual_adaptive_training,zhou2020in_situ_backprop_donn,pai2023in_situ_backprop_pnn,bandyopadhyay2024forward_only_pnn}. Details of the hardware-in-the-loop training architecture,
calibration procedure and SPSA update rule are provided in the Supplementary Information Chapter S4.5.

We first compare different single-task photonic architectures to identify the benefit of the
MPU-assisted optical core. As shown in Fig.~\ref{fig4}b, the MPU-based architecture
converges to a higher classification accuracy than the MZI-only and MPU-only baselines
under the same experimental workflow. The accuracy improvement of approximately
$6.8\%$ indicates that the inverse-designed MPU provides useful passive feature mixing,
rather than acting merely as a lossy optical interconnect. This result supports the central
design principle of the heterogeneous architecture: compact passive MPU operators provide
dense optical transformations, while the remaining trainable MZI components supply the
task-adaptive degrees of freedom needed for hardware calibration and learning.

We then evaluate simultaneous dual-task learning. The lower panel of Fig.~\ref{fig4}b shows
the training dynamics for two vowel classification tasks.Both task accuracies increase during the closed-loop update, while the training loss
decreases, indicating that the shared optical core can be tuned to support two related
classification objectives using measured hardware feedback.The final classification performance is summarized by the confusion matrices in
Fig.~\ref{fig4}c. For Task A, the experimental system achieves an inference accuracy of
$83.5\%$ on the test set. For Task B, the test accuracy reaches $80.9\%$. The diagonal
dominance of the confusion matrices indicates that the chip-level optical computation
preserves task-relevant class separability after hardware calibration. The remaining errors
mainly occur between acoustically similar vowel classes, consistent with the reduced
feature dimension and the small number of trainable physical parameters used in the
experiment.

Together, these results demonstrate a full task-level validation of the proposed
MPU-MZI photonic computing system. The experiment links inverse-designed passive
matrix operators, active electro-optic modulation, integrated photodetection and
hardware-aware optimization into a single closed-loop platform. Although the demonstrated
task is intentionally compact, the result verifies that the MPU can function as a useful
physical computing primitive inside an experimentally trained multi-task photonic neural
network.

\begin{figure}[H]
\centering
\includegraphics[width=1.0\textwidth]{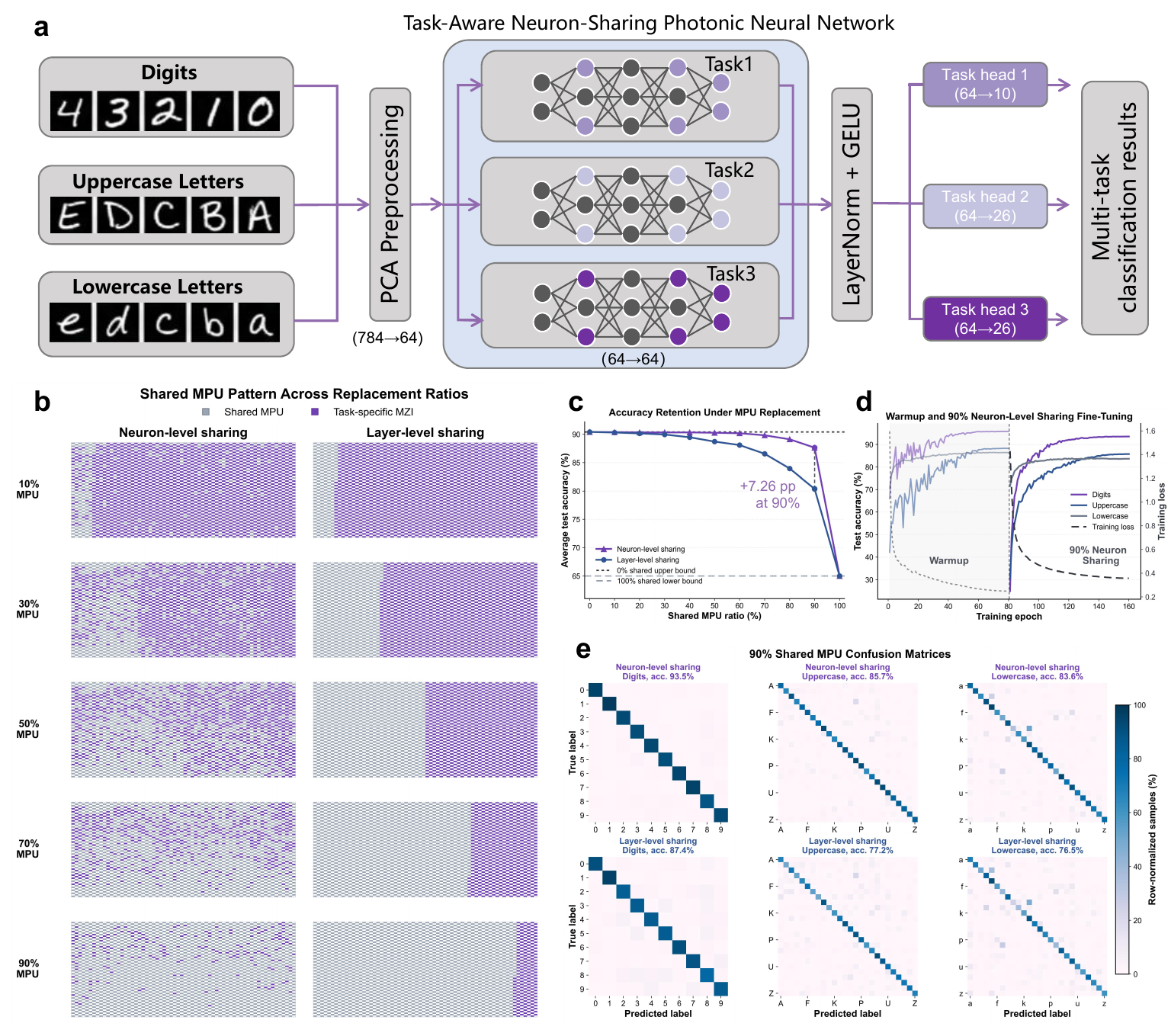}
\caption{
\textbf{Fine-grained MPU replacement in a multi-task photonic neural network.}
\textbf{a,} Overview of the proposed architecture. EMNIST samples from digit, uppercase-letter and lowercase-letter tasks are projected to 64 dimensions by fixed PCA and processed by a single coherent $64\times64$ Clements photonic mesh. The detected intensity is passed through a shared layer normalization and GELU module, followed by task-specific classification heads.
\textbf{b,} Shared MPU patterns at different replacement ratios. Grey denotes task-shared MPU units with common $2\times2$ transfer matrices across tasks, whereas purple denotes task-private MZI units with task-specific phase offsets.
\textbf{c,} Average test accuracy versus shared MPU ratio. Neuron-level sharing preserves higher accuracy than layer-level sharing under high replacement ratios, with a $7.26$ percentage-point gain at 90\% shared MPU replacement.
\textbf{d,} Training dynamics for the 90\% neuron-level condition. Dashed curves show the warmup stage before imposing the sharing mask, and solid curves show fine-tuning after the 90\% MPU pattern is fixed. The black curve reports training loss on the right axis.
\textbf{e,} Row-normalized confusion matrices at 90\% shared MPU replacement. Neuron-level sharing improves task-wise classification accuracy over layer-level sharing for all three tasks.
}
\label{fig5}
\end{figure}

\subsection{Large-scale EMNIST classification via fine-grained MPU replacement}\label{subsec:emnist}

To scale the optical neural network (ONN) for complex, larger multi-task visual classification while mitigating hardware footprint and control overhead, we investigate whether MPUs can replace a large fraction of reconfigurable Mach-Zehnder interferometer (MZI) neurons in a multi-task photonic neural network. Each inverse-designed MPU implements a compact $2\times2$ optical transfer matrix with a substantially smaller footprint and lower static power than a conventional tunable MZI, but becomes fixed after fabrication. Therefore, the key architectural question is not merely how many MZIs can be replaced, but where these passive MPU replacements should be placed to preserve task-specific optical adaptability.

We evaluate this question on a large-scale three-task EMNIST benchmark, consisting of digit, uppercase-letter, and lowercase-letter classification (Fig.~\ref{fig5}). The electronic-assisted photonic model uses fixed PCA preprocessing to compress each $28\times28$ input image from 784 to 64 dimensions, followed by a single coherent $64\times64$ Clements photonic mesh, square-law intensity readout,  LayerNorm and GELU activation, and task-specific electrical classification heads. This single-mesh setting directly matches the physical device question: each candidate replacement corresponds to one complete $2\times2$ MZI neuron in the Clements interferometer, rather than an isolated scalar weight or phase shifter. The task-specific heads prevent output-space competition from obscuring the optical replacement mechanism, so the comparison focuses on the placement strategy of shared passive MPU neurons inside the photonic mesh.

In this framework, each optical neuron can operate in one of two modes. If the neuron is assigned as a task-shared MPU, all three tasks use the same learned $2\times2$ transfer matrix, and task-specific phase offsets are disabled. In deployment, this common transformation is interpreted as a fixed passive MPU. If the neuron remains task-private, the corresponding MZI retains task-specific phase offsets and can locally adapt its optical transformation for each task. The shared MPU replacement ratio therefore quantifies the fraction of optical neurons whose task-specific reconfigurability is collapsed into common passive operators.

We compare two replacement strategies under exactly matched MPU budgets. The layer-level baseline adopts a coarse front-to-back hierarchical sharing prior, analogous to the common practice in deep multi-task visual networks: earlier layers are assumed to encode more task-shared feature transformations, whereas later layers are reserved for task-specific adaptation. Accordingly, this baseline replaces the front stages of the Clements mesh with shared MPUs and keeps the back stages as task-private MZIs. The proposed neuron-level strategy starts from the same front-to-back prior but refines it by identifying which individual neurons are more suitable for passive replacement. In this way, the neuron-level method does not use a lower replacement ratio or more task-private parameters; it only redistributes the same limited number of task-private MZIs to positions where task-specific optical adaptability is most valuable.

The accuracy curves reveal a clear separation between coarse and fine-grained replacement at high shared MPU ratios (Fig.~\ref{fig5}c). At low to moderate replacement ratios, both strategies maintain similar multi-task performance, indicating that a considerable fraction of the Clements mesh performs reusable or task-shared optical transformations. However, when the shared MPU ratio increases to 80-90\%, the few remaining task-private MZI neurons become especially critical. Coarse layer-level replacement confines these adaptive neurons to the back portion of the optical mesh, following the assumed hierarchy but lacking flexibility to account for local task-specific sensitivity. In contrast, neuron-level replacement distributes the limited task-private MZI budget across the mesh according to neuron-level commonness, preserving local adaptability at the most task-sensitive optical positions.

At $90\%$ shared MPU replacement, the neuron-level strategy achieves an average test accuracy of $87.64\%$, compared with $80.37\%$ for the layer-level baseline, corresponding to a $+7.26$ percentage-point improvement under the same passive replacement budget. Task-wise, neuron-level replacement reaches $93.5\%$ for digits, $85.7\%$ for uppercase letters, and $83.6\%$ for lowercase letters, whereas layer-level replacement yields $87.4\%$, $77.2\%$, and $76.5\%$, respectively. The confusion matrices further confirm that the performance gain is consistently observed across all three output spaces, rather than being dominated by a single task.

These results indicate that a large fraction of the optical mesh can be consolidated into shared passive MPU neurons without severe degradation, provided that the remaining reconfigurable MZI neurons are placed with sufficient spatial selectivity. The shared MPU neurons behave as reusable, task-agnostic optical transformations, while the sparse task-private MZIs supply the residual degrees of freedom required for task-specific adaptation. Fine-grained MPU replacement therefore provides a scalable route toward high-density multi-task photonic neural networks, retaining the compactness and low-control advantages of inverse-designed passive operators while preserving reconfigurability only where it is most functionally necessary.
%%%%%%% Discussion%%%%%%%%%%%%%%%%%%%%%%%%%%%%%%%%%%%%%%%%%%%%%%%
\section{Discussion}\label{sec:discussion}

This work demonstrates an inverse-designed meta processing unit (MPU) as a compact optical primitive for integrated photonic computing. By encoding local complex matrix operations into shallow-etched diffractive silicon structures, the MPU reduces the footprint and active tuning burden associated with conventional MZI-mesh-based implementations. The resulting architecture combines dense passive optical transformations with a limited number of active components for modulation, calibration and hardware-in-the-loop optimization, providing a practical route toward compact photonic neural-network processors.

The current implementation is based on a CMOS-compatible silicon photonic platform. Although this enables integration with MZMs, photodetectors and electrical control circuits, insertion loss and fabrication-induced scattering remain key constraints for large cascaded systems. Future implementations could benefit from lower-loss material platforms such as silicon nitride. Recent inverse-designed SiN devices have shown compact wavelength-, mode- and polarization-management functions with low-loss operation and fabrication robustness \cite{pita_ruiz2025sin_inverse_design}. Transferring MPU operators to such platforms could reduce passive propagation loss and improve the scalability of large multi-port photonic processors.

Beyond material optimization, the MPU concept can be extended to higher-dimensional optical computing. The present devices mainly operate in the spatial domain at a single wavelength. Future MPU modules could exploit space, wavelength, time and mode degrees of freedom simultaneously. Wavelength-division multiplexing could provide parallel matrix-operation channels, while mode multiplexing and time-domain modulation could further increase the effective computational throughput. Such high-dimensional MPUs may be particularly useful for multi-task or multimodal photonic computing, where different data streams can be processed in parallel within a shared photonic footprint.

The inverse-design framework can also be scaled by moving from monolithic device optimization to hierarchical design. For large input-output dimensions, directly optimizing a full electromagnetic region becomes computationally demanding. Because the shallow-etched MPU exhibits predominantly feed-forward propagation with weak back reflection, a large system can be approximated as a cascade of local scattering or transmission operators. This suggests a scalable route in which smaller sub-regions are inverse-designed separately or jointly through differentiable operator composition. Combining adjoint optimization, reduced-order electromagnetic models and surrogate solvers could further reduce the computational cost of generating large MPU libraries.

Finally, the hardware-in-the-loop experiments highlight the importance of hardware-aware training. Instead of relying only on ideal simulations, the measured optical response of the fabricated chip is included in the optimization loop, allowing the system to adapt to fabrication variations, coupling imbalance, detector mismatch and modulator nonidealities\cite{zheng2023dual_adaptive_training,zhou2020in_situ_backprop_donn,pai2023in_situ_backprop_pnn}. Future systems may combine this strategy with structured sparsity, task-conditioned gating and selective activation of MPU blocks, enabling different tasks to share a passive optical backbone while retaining task-specific optical flexibility.

In summary, the MPU is best viewed as a compact passive matrix-processing primitive that complements, rather than replaces, programmable photonic circuits. With further progress in low-loss materials, high-dimensional multiplexing and hierarchical inverse design, MPU-based architectures could provide a scalable foundation for dense, energy-efficient and hardware-adaptive photonic computing systems.

%%%%%%%%% Methods %%%%%%%%%%%%%%%%%%%%%%%%%%%%%%%%%%%%%%%%%%%%%%%%%%%%%%%%%% 
\section{Methods}\label{sec:methods}

\subsection{Electromagnetic inverse-design formulation}\label{subsec:math}
The electromagnetic response of the proposed MPU was obtained by solving the frequency-domain Maxwell equation
\begin{equation}
\nabla \times \mu_0^{-1}\nabla \times \mathbf{E}(\mathbf{r})
-
\omega^2 \varepsilon(\mathbf{r})\mathbf{E}(\mathbf{r})
=
i\omega \mathbf{J}(\mathbf{r}).
\end{equation}
After spatial discretization, the problem can be written as a sparse linear system
\begin{equation}
\mathbf{A}(\boldsymbol{\varepsilon})\mathbf{e}=\mathbf{b},
\end{equation}
where $\mathbf{A}(\boldsymbol{\varepsilon})$ is the discretized Maxwell operator, $\mathbf{e}$ is the vectorized electromagnetic field, and $\mathbf{b}$ denotes the input excitation. Formally, the solution is $\mathbf{e}=\mathbf{A}^{-1}\mathbf{b}$, meaning that electromagnetic simulation evaluates the action of the inverse Maxwell operator on the source term. The inverse matrix is not explicitly constructed; instead, the sparse Maxwell system is solved numerically.

The material distribution in the design region was parameterized by design variables $\mathbf{p}$, with $\boldsymbol{\varepsilon}=\boldsymbol{\varepsilon}(\mathbf{p})$. The inverse design problem was formulated as maximizing a differentiable figure of merit $\mathcal{F}(\mathbf{e},\mathbf{p})$ subject to Maxwell's equations. To efficiently compute gradients in the high-dimensional design space, we used the adjoint-variable method. For a design variable $p_i$, differentiating the forward system gives
\begin{equation}
\mathbf{A}\frac{\partial \mathbf{e}}{\partial p_i}
=
-\frac{\partial \mathbf{A}}{\partial p_i}\mathbf{e}.
\end{equation}
Introducing the adjoint field $\boldsymbol{\lambda}$ satisfying
\begin{equation}
\mathbf{A}^{\dagger}\boldsymbol{\lambda}
=
\frac{\partial \mathcal{F}}{\partial \mathbf{e}^{*}},
\end{equation}
the gradient can be evaluated as
\begin{equation}
\frac{\partial \mathcal{F}}{\partial p_i}
=
-2\mathrm{Re}
\left[
\boldsymbol{\lambda}^{\dagger}
\frac{\partial \mathbf{A}}{\partial p_i}
\mathbf{e}
\right].
\end{equation}
Thus, after one forward simulation and one adjoint simulation, the sensitivities of all design variables can be obtained from local overlap integrals between the forward and adjoint fields.

The shallow-etched geometry used in this work exhibits weak backward scattering, with more than 95\% of the optical power propagating in the forward direction. This near-unidirectional propagation enables a scalable hierarchical interpretation of the inverse-design problem. When the design region is partitioned along the propagation direction, the global electromagnetic response can be approximated as a cascade of local propagation operators,
\begin{equation}
\mathbf{e}_{\mathrm{out}}
\approx
\mathbf{P}_{L}\mathbf{P}_{L-1}\cdots \mathbf{P}_{1}\mathbf{e}_{\mathrm{in}},
\end{equation}
where $\mathbf{P}_{l}$ denotes the effective local transmission operator of the $l$-th sub-region. In this representation, the global response can be interpreted as a composition of smaller
local transmission operators, suggesting a scalable route based on local electromagnetic
subproblems rather than a monolithic design domain.
\subsection{Experimental apparatus and hardware implementation}

Optical and electrical measurements were performed on a fiber-coupled silicon photonic
test platform. A tunable continuous-wave laser (Ceyear 6317A, 1480-1640 nm) provided
the optical input. When required, an erbium-doped fiber amplifier (Thorlabs EDFA300S(X))
was used to compensate coupling and chip insertion losses. For wavelength-dependent
characterization, the laser wavelength was swept and the output powers were recorded by a
high-speed optical power meter (Ceyear 6337F).

For passive scattering-matrix measurements, light was coupled into selected input ports
through a polarization controller and a lensed fiber, while the corresponding output powers
were measured to reconstruct the transmission matrix. For active photonic computing
experiments, the on-chip Mach-Zehnder modulators and phase-tuning electrodes were
driven by a multi-channel programmable voltage source (T1-MS128-12CV, Jiaxing
Time-transfer Optoelectronics Co., Ltd.). Photocurrents from the integrated photodetectors
were measured using a picoammeter (Keithley 6485) and transferred to a host computer for
normalization, loss evaluation and hardware-in-the-loop parameter updates.

The active silicon photonic chips were fabricated through the AMF multi-project-wafer
service and packaged by Chongqing United Microelectronics Center (CUMEC), including
die mounting, wire bonding and PCB-level electrical fan-out. Although the experiments in
this work used quasi-static or DC control for stable calibration and closed-loop training, the
chip and package were designed with high-frequency electrical interfaces compatible with
GHz-bandwidth MZM driving. The inverse-designed MPU patterns were generated using the
open-source Stanford SPINS framework, with electromagnetic optimization performed on a
GPU server equipped with four NVIDIA RTX 3090 GPUs.

\subsection{Hardware-in-the-loop multi-task photonic training}

To validate task-level optical computing on the fabricated chip, we implemented a
hardware-in-the-loop multi-task training system that combines digital preprocessing,
calibrated electro-optic encoding, on-chip photonic computation and digital readout
calibration. The hybrid model contains a lightweight digital pre-processing network
(PreNet), an electro-optic input interface based on Mach-Zehnder modulators (MZMs), an
MPU-MZI photonic computing core, integrated photodetector readout and a digital
classification backend.

For each input sample, the raw feature vector is first compressed by the PreNet into a
four-dimensional representation,
\begin{equation}
    \mathbf{z}
    =
    \sigma
    \left(
    \mathbf{W}_{\mathrm{pre}}\mathbf{x}
    +
    \mathbf{b}_{\mathrm{pre}}
    \right),
\end{equation}
where $\sigma(\cdot)$ denotes the sigmoid function. The resulting normalized features are
mapped to physical MZM driving voltages through calibrated channel-wise voltage windows,
\begin{equation}
    V_i
    =
    V_{\mathrm{LO},i}
    +
    \alpha_i
    \left(
    V_{\mathrm{HI},i}
    -
    V_{\mathrm{LO},i}
    \right)
    z_i .
\end{equation}
Here, $V_{\mathrm{LO},i}$ and $V_{\mathrm{HI},i}$ are obtained from the measured linear
operating region of the $i$-th MZM, and $\alpha_i$ is a trainable input-scaling coefficient.
Before training, the MZM voltage-transmission curves are swept experimentally to identify
a monotonic quasi-linear region, so that voltage updates produce predictable optical
amplitude modulation.

The optical signals then propagate through the MPU-MZI photonic core, whose physical
configuration is controlled by a set of trainable phase parameters $\boldsymbol{\theta}$.
The output optical intensities are converted to photocurrents by the integrated
photodetectors. To compensate channel-dependent detector responsivity, path loss and
static readout imbalance, the measured photocurrent vector is calibrated as
\begin{equation}
    \mathbf{I}_{\mathrm{cal}}
    =
    \boldsymbol{\beta}
    \odot
    \mathbf{I}_{\mathrm{raw}},
\end{equation}
where $\boldsymbol{\beta}$ is a trainable readout-scaling vector and $\odot$ denotes
element-wise multiplication. The calibrated readout is subsequently normalized and used by
the digital backend to compute the classification output and the task loss.

The complete trainable parameter set is
\begin{equation}
    \boldsymbol{\Theta}
    =
    \{
    \mathbf{W}_{\mathrm{pre}},
    \mathbf{b}_{\mathrm{pre}},
    \boldsymbol{\alpha},
    \boldsymbol{\theta},
    \boldsymbol{\beta}
    \}.
\end{equation}
In our implementation, this corresponds to 64 trainable parameters: 44 parameters in the
digital PreNet, 4 MZM input-scaling coefficients, 12 physical phase settings in the photonic
core and 4 readout calibration factors.
 
Because the fabricated photonic chip contains unknown hardware nonidealities, including
fabrication variations, coupling imbalance, detector responsivity mismatch and thermal
drift, we optimize the hybrid model directly using hardware measurements rather than
relying only on a simulated differentiable model. We use simultaneous perturbation
stochastic approximation (SPSA), a gradient-free stochastic optimization method that
requires only two hardware evaluations per iteration regardless of the number of trainable
parameters\cite{spall1992multivariate_spsa,bandyopadhyay2024forward_only_pnn}. At iteration $k$, a random Bernoulli perturbation vector
$\boldsymbol{\Delta}_k$ is sampled, and two perturbed parameter settings are applied to the
hardware,
\begin{equation}
    \boldsymbol{\Theta}_k^{+}
    =
    \boldsymbol{\Theta}_k
    +
    c_k\boldsymbol{\Delta}_k,
    \qquad
    \boldsymbol{\Theta}_k^{-}
    =
    \boldsymbol{\Theta}_k
    -
    c_k\boldsymbol{\Delta}_k .
\end{equation}
The corresponding hardware-measured losses, $L_k^{+}$ and $L_k^{-}$, are obtained from
the photocurrent readout and the digital classification backend. The stochastic gradient is
estimated as
\begin{equation}
    \widehat{\mathbf{g}}_k
    =
    \frac{
    L_k^{+}-L_k^{-}
    }{
    2c_k
    }
    \boldsymbol{\Delta}_k^{-1},
\end{equation}
where the reciprocal is taken element-wise. Since each entry of
$\boldsymbol{\Delta}_k$ is $\pm1$, this operation is equivalent to multiplying by
$\boldsymbol{\Delta}_k$. The parameters are then updated by
\begin{equation}
    \boldsymbol{\Theta}_{k+1}
    =
    \boldsymbol{\Theta}_k
    -
    a_k
    \widehat{\mathbf{g}}_k ,
\end{equation}
where $a_k$ is the learning rate. After each update, the MZM driving voltages are clipped
to their calibrated linear operating regions.

This closed-loop procedure allows the digital parameters, electro-optic input scaling,
physical photonic phase settings and readout calibration factors to be jointly adapted to
the actual chip response. Therefore, the trained model automatically incorporates hardware
imperfections into the optimization process, enabling robust multi-task classification with
the fabricated MPU photonic processor. Detailed parameter allocation, calibration
procedures and the full SPSA-based training workflow are provided in the Supplementary Information Chapter S4.

\subsection{Fine-grained MPU replacement strategy for EMNIST simulation}

To evaluate the scalability of the proposed MPU-based photonic computing architecture, we
performed a multi-task EMNIST simulation including digit, uppercase-letter and
lowercase-letter classification. Each $28\times28$ image was flattened and projected to a
64-dimensional optical input vector using a fixed PCA layer. The optical core was modeled
as a coherent $64\times64$ Clements mesh followed by square-law detection, shared
normalization and activation layers, and task-specific electronic classification heads. The
task-specific heads isolate the three label spaces, so that the comparison mainly reflects the
optical-core sharing strategy.

Each local $2\times2$ unit in the Clements mesh was treated as a replaceable optical neuron.
In the task-private reference model, each unit is implemented by a programmable MZI with
task-specific phase offsets. For task $t$ and optical unit $i$, the effective phase settings are
\begin{equation}
    \theta_{t,i}^{\mathrm{eff}}
    =
    \theta_i+\Delta\theta_{t,i},
    \qquad
    \phi_{t,i}^{\mathrm{eff}}
    =
    \phi_i+\Delta\phi_{t,i}.
\end{equation}
A binary mask determines whether the unit is retained as a task-private MZI or replaced by
a shared passive MPU,
\begin{equation}
    M_i =
    \begin{cases}
    1, & \mathrm{shared\ MPU},\\
    0, & \mathrm{task\text{-}private\ MZI}.
    \end{cases}
\end{equation}
When $M_i=1$, the task-specific offsets are disabled and all tasks share the same local
$2\times2$ operator. When $M_i=0$, the offsets remain enabled and the unit preserves
task-specific optical tunability. The MPU replacement ratio is therefore
\begin{equation}
    r =
    \frac{1}{N}
    \sum_{i=1}^{N} M_i ,
\end{equation}
where $N$ is the total number of replaceable optical units in the mesh.

We used a two-stage training protocol. First, a warmup model was trained with
$M_i=0$ for all optical units, providing a fully task-private reference. Second, for each target
replacement ratio, a fixed binary replacement mask was imposed and the model was
fine-tuned from the warmup checkpoint. This protocol allows different sharing strategies to
be compared under the same passive-MPU budget.

As a coarse baseline, we used a stage-level replacement rule that replaces Clements stages
sequentially from the input side until the target replacement ratio is reached. To obtain a
more selective placement of shared MPUs, we further used a neuron-level replacement
strategy based on a Fisher-style damage score. For each optical unit, the task-specific
phases are first collapsed to a common circular mean, denoted by
$\bar{\theta}_i$ and $\bar{\phi}_i$. The phase mismatch caused by this collapse is measured as
\begin{equation}
    d_{t,i}
    =
    \frac{1}{2}
    \left[
    \mathrm{wrap}
    \left(
    \theta_{t,i}^{\mathrm{eff}}-\bar{\theta}_i
    \right)^2
    +
    \mathrm{wrap}
    \left(
    \phi_{t,i}^{\mathrm{eff}}-\bar{\phi}_i
    \right)^2
    \right],
\end{equation}
where $\mathrm{wrap}(x)=\mathrm{atan2}(\sin x,\cos x)$ accounts for phase periodicity. The
importance of the unit is estimated by a diagonal Fisher proxy,
\begin{equation}
    F_{t,i}
    =
    \left(
    \frac{\partial \mathcal{L}_t}{\partial \theta_i}
    \right)^2
    +
    \left(
    \frac{\partial \mathcal{L}_t}{\partial \phi_i}
    \right)^2 .
\end{equation}
The estimated replacement damage is then
\begin{equation}
    D_i
    =
    \frac{1}{T}
    \sum_{t=1}^{T}
    F_{t,i} d_{t,i},
\end{equation}
where $T=3$ is the number of tasks. Optical units with low replacement damage and high
cross-task phase consistency are preferentially replaced by shared MPUs, whereas
task-sensitive units are retained as task-private MZIs.

For a fair comparison, the neuron-level mask preserves the same number of shared units as
the stage-level baseline,
\begin{equation}
    \sum_{i=1}^{N}
    M_i^{\mathrm{neuron}}
    =
    \sum_{i=1}^{N}
    M_i^{\mathrm{stage}}
    \approx rN .
\end{equation}
Thus, the two strategies have the same replacement ratio and the same number of remaining
task-private MZIs; performance differences arise only from the spatial placement of shared
MPUs. This fine-grained strategy preserves task-specific tunability at high-impact optical
locations while replacing most of the mesh with compact passive MPU primitives. Detailed
mask construction, training hyperparameters and visualizations of the learned phase and
energy-flow distributions are provided in the Supplementary Information Chapter S5.
\backmatter

\section*{Author contributions}
Chu Wu: Conceptualization, methodology, numerical implementation, inverse design, simulations, experiment, data analysis, visualization, writing—original draft, writing—review and editing.
Zeyu Cai: Experiment setup, optical measurements, data acquisition, result validation.
Songtao Yang: Scientific discussion, result analysis, manuscript feedback.
Yinan Zhao, Haiou Zhang: Scientific discussion.
Ruoyu Shen: Device microfabrication, chip processing, experimental support.
Wei Chu: Fabrication guidance, device testing, process optimization.
Xing Lin: Supervision, project administration and initiation, funding acquisition, conceptual guidance, methodology and interpretation, writing—review and editing.

\section*{Declarations}
\begin{itemize}
\item Funding: This work is supported by the National Key Research and Development Program of China (No. BNR2026RC01008, No. 2021ZD0109902), National Natural Science Foundation of China (No. 62275139).
\item Conflict of interest: The authors declare no conflict of interest.
\item Material and Code availability: Data presented in this publication is available on GitHub with the following link: https://github.com/THPCILab/MPU. The codes used in the current study are available from the corresponding authors upon reasonable request.
\end{itemize}

% =====================================================================
% Supplementary Information (integrated into the arXiv source)
% =====================================================================
\clearpage
\setcounter{section}{0}
\setcounter{subsection}{0}
\setcounter{figure}{0}
\setcounter{table}{0}
\setcounter{equation}{0}
\renewcommand{\thesection}{S\arabic{section}}
\renewcommand{\thesubsection}{\thesection.\arabic{subsection}}
\renewcommand{\thefigure}{S\arabic{figure}}
\renewcommand{\thetable}{S\arabic{table}}
\renewcommand{\theequation}{S\arabic{equation}}
\makeatletter
\@ifundefined{theHfigure}{}{\renewcommand{\theHfigure}{S\arabic{figure}}}
\@ifundefined{theHsection}{}{\renewcommand{\theHsection}{S\arabic{section}}}
\@ifundefined{theHsubsection}{}{\renewcommand{\theHsubsection}{S\arabic{section}.\arabic{subsection}}}
\@ifundefined{theHtable}{}{\renewcommand{\theHtable}{S\arabic{table}}}
\@ifundefined{theHequation}{}{\renewcommand{\theHequation}{S\arabic{equation}}}
\makeatother
\begin{center}
{\Large\bfseries Supplementary Information for\par}
\vspace{0.45em}
{\large\bfseries Inverse-designed meta processing units for multi-task near-field photonic computing\par}
\vspace{0.75em}
{\normalsize Chu Wu, Zeyu Cai, Songtao Yang, Ruoyu Shen, Yinan Zhao, Haiou Zhang, Wei Chu and Xing Lin\par}
\end{center}
\vspace{0.75em}

\section{Inverse-designed Meta Processing Unit Device Library}

\subsection{Inverse Design Methodology}

The inverse-designed meta processing unit (MPU) was optimized to realize a prescribed
$2\times2$ complex scattering matrix between two input and two output waveguide modes.
The device response was evaluated by frequency-domain electromagnetic simulations at
$\lambda=1550~\mathrm{nm}$. In the discretized finite-difference frequency-domain (FDFD)
formulation, Maxwell's equations can be written as a sparse linear system
\begin{equation}
    \mathbf{A}(\boldsymbol{\varepsilon})\mathbf{e}=\mathbf{b},
    \label{eq:supp_maxwell_linear}
\end{equation}
where $\mathbf{A}(\boldsymbol{\varepsilon})$ is the permittivity-dependent Maxwell operator,
$\mathbf{e}$ is the vectorized electric field, and $\mathbf{b}$ represents the waveguide-mode
excitation. Formally, the field solution is
\begin{equation}
    \mathbf{e}=\mathbf{A}^{-1}(\boldsymbol{\varepsilon})\mathbf{b},
    \label{eq:supp_maxwell_inverse}
\end{equation}
which indicates that the electromagnetic solver evaluates the action of the inverse Maxwell
operator on the excitation source. In practice, the inverse matrix is never explicitly formed;
instead, Eq.~\eqref{eq:supp_maxwell_linear} is solved numerically by an iterative Maxwell
solver.

The design region is represented by a pixelated continuous variable
$\mathbf{x}\in[0,1]^{N}$, where each pixel interpolates between the shallow-etched and
unetched silicon regions. To gradually enforce fabricable binary structures, the physical
design variable is projected through a sigmoid continuation function,
\begin{equation}
    \rho(\mathbf{x};\beta)
    =
    \sigma\!\left[\beta(2\mathbf{x}-1)\right],
    \label{eq:supp_sigmoid}
\end{equation}
where $\sigma(\cdot)$ is the sigmoid function and $\beta$ controls the sharpness of the
projection. During optimization, $\beta$ is progressively increased so that the optimized
permittivity distribution evolves from a smooth continuous pattern to a nearly binary
geometry.

For a target matrix
\begin{equation}
    \mathbf{S}^{\mathrm{targ}}
    =
    \begin{bmatrix}
    s_{11}^{\mathrm{targ}} & s_{12}^{\mathrm{targ}} \\
    s_{21}^{\mathrm{targ}} & s_{22}^{\mathrm{targ}}
    \end{bmatrix},
    \label{eq:supp_target_matrix}
\end{equation}
the two columns of $\mathbf{S}^{\mathrm{targ}}$ are optimized by launching the fundamental
waveguide mode from the upper and lower input ports, respectively. For excitation from input
port 1, the complex forward mode-overlap coefficients at the two output ports are denoted
as $s_{11}^{\mathrm{sim}}$ and $s_{21}^{\mathrm{sim}}$, while the backward overlaps at the two
input-side ports are denoted as $r_{11}^{\mathrm{sim}}$ and $r_{21}^{\mathrm{sim}}$. The
corresponding loss is defined as
\begin{equation}
    \mathcal{L}_{1}
    =
    \left|
    \sqrt{\eta}\,s_{11}^{\mathrm{targ}}-s_{11}^{\mathrm{sim}}
    \right|^{2}
    +
    \left|
    \sqrt{\eta}\,s_{21}^{\mathrm{targ}}-s_{21}^{\mathrm{sim}}
    \right|^{2}
    +
    \left|r_{11}^{\mathrm{sim}}\right|^{2}
    +
    \left|r_{21}^{\mathrm{sim}}\right|^{2}.
    \label{eq:supp_loss_port1}
\end{equation}
Similarly, for excitation from input port 2, the simulated forward overlaps are
$s_{12}^{\mathrm{sim}}$ and $s_{22}^{\mathrm{sim}}$, and the backward overlaps are
$r_{12}^{\mathrm{sim}}$ and $r_{22}^{\mathrm{sim}}$. The second-port loss is
\begin{equation}
    \mathcal{L}_{2}
    =
    \left|
    \sqrt{\eta}\,s_{12}^{\mathrm{targ}}-s_{12}^{\mathrm{sim}}
    \right|^{2}
    +
    \left|
    \sqrt{\eta}\,s_{22}^{\mathrm{targ}}-s_{22}^{\mathrm{sim}}
    \right|^{2}
    +
    \left|r_{12}^{\mathrm{sim}}\right|^{2}
    +
    \left|r_{22}^{\mathrm{sim}}\right|^{2}.
    \label{eq:supp_loss_port2}
\end{equation}
Here, $\eta$ is the target transmission efficiency. The factor $\sqrt{\eta}$ scales the target
complex amplitudes so that the optimized device is encouraged to match both the desired
complex scattering relation and the specified insertion-loss budget. The backward-overlap
terms explicitly penalize reflected or backward-propagating power, thereby suppressing
spurious feedback and enforcing the intended feed-forward optical response.

The total optimization objective is
\begin{equation}
    \mathcal{L}_{\mathrm{tot}}
    =
    \mathcal{L}_{1}
    +
    \mathcal{L}_{2}.
    \label{eq:supp_total_loss}
\end{equation}
This objective directly optimizes the complex scattering response of the two-port device
rather than only its output intensity. Both amplitude and phase errors are included through
the complex-valued overlap mismatch, while reflection leakage is suppressed through the
backward-mode penalty. In this way, the designed MPU implements a physically constrained
matrix operator with controlled transmission efficiency and reduced back reflection.

The gradient of the objective with respect to all design pixels is computed using the adjoint
method. For each excitation condition, one forward simulation obtains the physical field, and
one adjoint simulation provides the sensitivity of the objective to the permittivity distribution.
Equivalently, for a design parameter $p_i$, differentiating
Eq.~\eqref{eq:supp_maxwell_linear} gives
\begin{equation}
    \mathbf{A}\frac{\partial \mathbf{e}}{\partial p_i}
    =
    -
    \frac{\partial \mathbf{A}}{\partial p_i}
    \mathbf{e}.
    \label{eq:supp_field_derivative}
\end{equation}
Introducing the adjoint field $\boldsymbol{\lambda}$ satisfying
\begin{equation}
    \mathbf{A}^{\dagger}\boldsymbol{\lambda}
    =
    \frac{\partial \mathcal{L}_{\mathrm{tot}}}{\partial \mathbf{e}^{*}},
    \label{eq:supp_adjoint_system}
\end{equation}
the gradient can be evaluated from the local overlap between the forward and adjoint fields,
\begin{equation}
    \frac{\partial \mathcal{L}_{\mathrm{tot}}}{\partial p_i}
    =
    -2\mathrm{Re}
    \left[
    \boldsymbol{\lambda}^{\dagger}
    \frac{\partial \mathbf{A}}{\partial p_i}
    \mathbf{e}
    \right].
    \label{eq:supp_adjoint_gradient}
\end{equation}
Thus, the sensitivities of all design pixels are obtained without performing a separate
finite-difference simulation for each pixel, enabling efficient optimization of high-dimensional
subwavelength design regions.

In the implementation, the MPU design region is discretized into square pixels, embedded in
a silicon-on-insulator waveguide platform, and surrounded by perfectly matched layers to
absorb outgoing radiation. The upper and lower input waveguides are excited separately by
the fundamental guided mode, and complex waveguide-mode overlaps are monitored at both
forward output ports and backward input-side ports. The optimization is performed using
L-BFGS-B with a continuation schedule on the sigmoid sharpness parameter. Specifically, the
projection factor is increased sequentially to refine the design from a continuous permittivity
profile to a near-binary structure, while each continuation stage minimizes
$\mathcal{L}_{\mathrm{tot}}$. This staged procedure improves convergence stability, suppresses
unphysical grayscale features, and balances matrix fidelity, insertion loss, and reflection
suppression.

Performance of the designed devices is quantified via a comprehensive evaluation framework: complex mean squared error (MSE) for full $\bm{S}$ matrix fitting (retaining amplitude/phase information), matrix fidelity (a normalized structural similarity metric for complex matrices), power efficiency (normalized total power transmission), and bit precision (normalized maximum element error in logarithmic scale). This normalized fidelity evaluates matrix-structure similarity and is therefore reported together with transmission efficiency. A power-normalized evaluation further corrects for total power mismatch between target and simulated matrices, enabling fair comparison of element-wise relative errors while avoiding division by zero. All optimizations are performed over the pixelated permittivity distribution, with the adjoint method (Eq.2) enabling efficient gradient computation to minimize the adaptive objective function (Eq. 3), and the four-stage workflow (Algorithm 1) balancing amplitude/phase matching, reflection suppression, and fabrication feasibility via sigmoid-driven binarization.

\begin{figure}[H]
\centering
\includegraphics[width=0.6\textwidth]{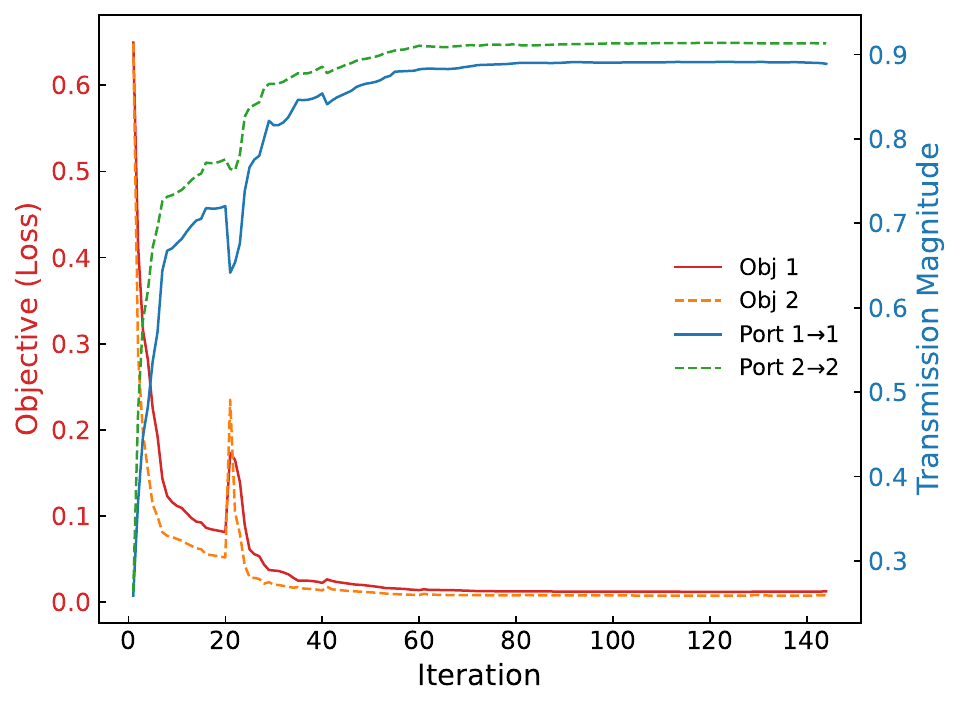}
\caption{\textbf{Convergence dynamics of the inverse-designed optical bar-state operator. Evolution of the multi-objective loss function and transmission magnitudes during the adjoint optimization process of a 2×2 routing device.} The left vertical axis (red lines) illustrates the progressive minimization of the Mean Squared Error (MSE) objective function for two individual excitation conditions (Obj 1, solid line; Obj 2, dashed line). The right vertical axis (blue lines) tracks the corresponding transmission magnitudes for the target optical routing paths (Port 1→1, solid line; Port 2→2, dashed line). Over the designated optimization iterations, the multiobjective loss symmetrically converges towards its minimum, yielding final transmission magnitudes around 0.89 ($\sim$79 $\%$ coupling energy efficiency) for both target channels.}
\label{fig:s1}
\end{figure}

\subsection{Parametric scaling and efficiency--fidelity trade-off}
\begin{figure}[ht]
\centering
\includegraphics[width=0.9\textwidth]{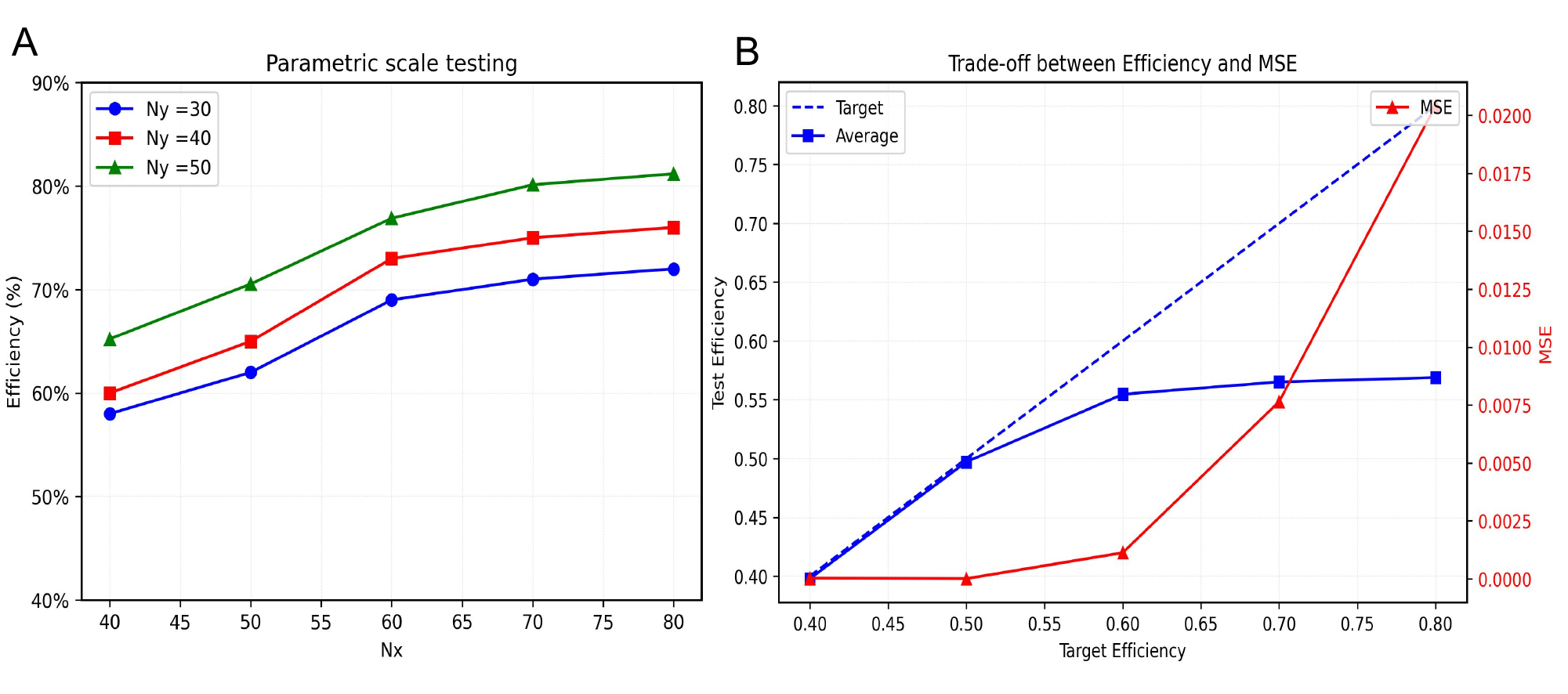}
\caption{\textbf{Parametric spatial scanning and performance trade-off analysis in inverse-designed optical devices.} 
(a) Physical footprint scaling: Illustrates the impact of the design region's grid dimensions (length scale Nx and width scale Ny) on the output efficiency. 
(b) The trade-off between accuracy and transmission efficiency: Depicts the relationship across the target efficiency constraints, the tested average transmission efficiency, and the Mean Squared Error (MSE).}
\label{fig:s2}
\end{figure}

To analyze the design-size dependence and the insertion-loss--fidelity trade-off in a
controlled setting, we selected a representative bar-state operator as the inverse-design
target,
\begin{equation}
    \mathbf{S}^{\mathrm{target}}
    =
    \begin{bmatrix}
    1 & 0\\
    0 & 1
    \end{bmatrix}.
\end{equation}
This identity-like target corresponds to routing the upper input to the upper output and the
lower input to the lower output, and therefore provides a convenient benchmark for
evaluating the achievable transmission efficiency and the required spatial degrees of
freedom without additional matrix-mixing complexity.

As shown in Fig.~S2a, increasing the longitudinal size $N_x$ consistently improves the
transmission efficiency for all tested transverse apertures. For example, when $N_y=30$,
the efficiency increases from approximately $58\%$ at $N_x=40$ to about $72\%$ at
$N_x=80$. With a larger transverse aperture of $N_y=50$, the optimized device reaches
above $80\%$ transmission efficiency. This confirms that enlarging the design region
provides more spatial degrees of freedom for wavefront evolution and suppresses
non-useful scattering loss.

Figure~S2b further reveals the trade-off between insertion loss and complex-matrix
fidelity under this bar-state target. As the target efficiency increases from 0.4 to 0.8, the
tested average transmission rises from approximately $40\%$ to $57\%$. However, the
matrix MSE remains low only in the moderate-efficiency regime and increases sharply when
the target efficiency is forced to 0.8. This indicates that, even for the relatively simple
identity-like operator, maximizing transmission and minimizing complex matrix error cannot
be achieved independently within an ultra-compact footprint. A purely fidelity-driven
objective may yield an apparently accurate but low-transmission operator, whereas an
overly aggressive transmission constraint can distort the desired scattering response.

Therefore, the target-efficiency term and the back-reflection penalty are explicitly included
in the inverse-design objective. These terms guide the optimizer toward a practical Pareto
region between low insertion loss and high matrix fidelity. In this representative test, the
single-device transmission efficiency can reach approximately $80\%$, whereas
conventional ultra-compact metaline-type diffractive operators typically exhibit only
percent- to sub-percent-level transmission. This corresponds to an improvement of one to
nearly three orders of magnitude in single-device transmission efficiency, depending on the
specific metaline reference. The result should be interpreted as a controlled benchmark for
evaluating parameter scaling and insertion-loss limits, rather than as the average efficiency
of all possible complex matrix operators in the device library.

\subsection{Quantized unitary-device library and non-unitary operator extension}

We next verify the MPU device library on a discretized unitary operator space. The
ground-truth operators are generated from the standard $2\times2$ MZI transfer matrix
parameterized by two phase-control variables $(\theta,\phi)$,
\begin{equation}
    \mathbf{U}(\theta,\phi)
    =
    i e^{i\theta/2}
    \begin{bmatrix}
    e^{i\phi}\sin(\theta/2) & e^{i\phi}\cos(\theta/2)\\
    \cos(\theta/2) & -\sin(\theta/2)
    \end{bmatrix}.
    \label{eq:supp_mzi_unitary}
\end{equation}
To construct a finite device library compatible with digitally controlled photonic hardware,
both $\theta$ and $\phi$ are quantized using a 3-bit grid. This produces an $8\times8$
set of target unitary transformations, corresponding to 64 representative MZI-equivalent
operators.

For each quantized $(\theta,\phi)$ pair, an individual shallow-etched MPU is
inverse-designed to reproduce the corresponding complex scattering matrix. The resulting
device library is then validated by full 3D FDTD simulations. As shown in Fig.~S3, the
theoretical complex output points generated from the target matrices are compared with
the simulated complex points extracted from the FDTD near-field responses of the
inverse-designed devices. The close overlap between the simulated and target points
across all 64 configurations confirms that the MPU library can faithfully reproduce the
3-bit quantized unitary operator space. This validates the consistency of the inverse-designed
primitive library for digitally selected optical matrix operations and cascaded photonic
computing.

Although the above library is constructed from 3-bit quantized MZI parameters, the MPU
design framework is not intrinsically restricted to unitary transformations. Under a fixed
target-efficiency constraint, the accessible operator space can be extended to general lossy
complex matrices. To demonstrate this capability, we parameterize a general $2\times2$
complex matrix using an SVD-inspired form,
\begin{equation}
    \mathbf{T}
    =
    \mathbf{U}_{\mathrm{L}}
    \begin{bmatrix}
    s_1 & 0\\
    0 & s_2
    \end{bmatrix}
    \mathbf{U}_{\mathrm{R}},
    \label{eq:supp_general_matrix}
\end{equation}
where $\mathbf{U}_{\mathrm{L}}$ and $\mathbf{U}_{\mathrm{R}}$ are unitary matrices and
$s_1,s_2$ are singular-value-like amplitude scaling factors. This representation extends
the target space from energy-conserving unitary transformations to general non-unitary
operators with independent amplitude and phase degrees of freedom.

As a representative validation, we generated 100 arbitrary target matrices using
Eq.~\eqref{eq:supp_general_matrix} and performed inverse design under a fixed target
transmission efficiency of $25\%$. The optimized devices achieved an average matrix
fidelity of $97.76\%$, indicating that high-fidelity MPU synthesis remains feasible beyond
the unitary-matrix subspace. Therefore, the 64-device quantized MZI library should be
viewed as a digitally addressable subset of a broader inverse-designable operator space,
rather than as the fundamental limit of the MPU. This extension substantially enriches the
available solution space for photonic neural-network accelerators, where practical weight
matrices are generally non-unitary and require both amplitude and phase modulation.

\begin{figure}[ht]
\centering
\includegraphics[width=1.0\textwidth]{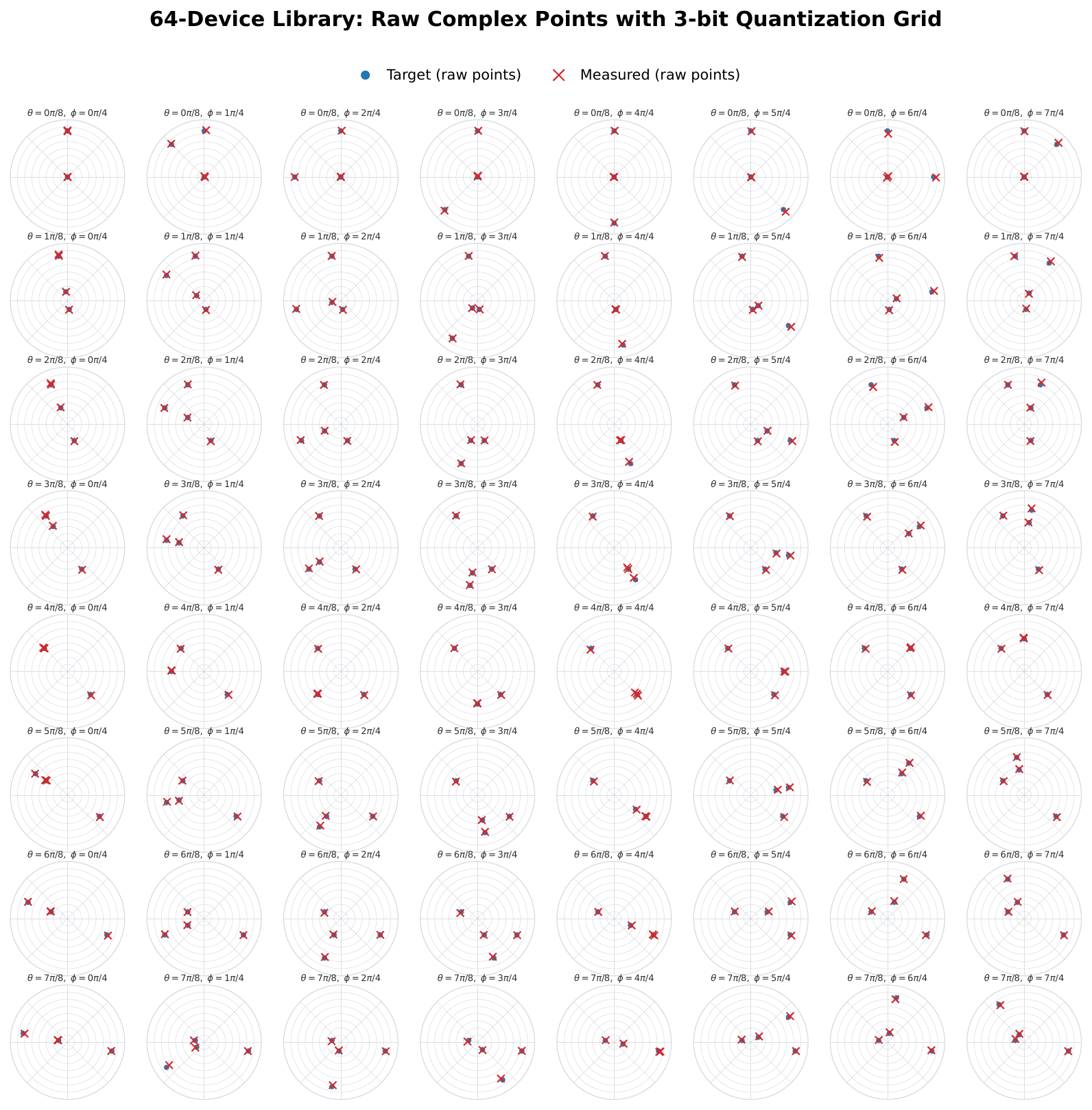}
\caption{\textbf{ Systematic validation of the MPU photonic operator library via 3D FDTD simulation versus theoretical ground-truth state comparison.}}
\label{fig:s3}
\end{figure}

\subsection{Matrix-operator bit precision.}
We further quantify the numerical precision of the inverse-designed operator library by
converting the complex-matrix reconstruction error into an equivalent bit precision. It is
important to distinguish this metric from the 3-bit quantization used to generate the target
MZI control grid. The 3-bit grid defines the discrete set of target operators, whereas the
operator bit precision evaluates how accurately each inverse-designed MPU reproduces its
assigned complex matrix.

For a target matrix $\mathbf{T}$ and the corresponding simulated matrix
$\mathbf{S}^{\mathrm{sim}}$, we first evaluate the loss-normalized matrix error to remove
the trivial effect of global insertion loss,
\begin{equation}
    \epsilon_{\mathrm{rms}}
    =
    \frac{
    \left\|
    \widetilde{\mathbf{S}}^{\mathrm{sim}}-\mathbf{T}
    \right\|_{\mathrm{F}}
    }{
    \left\|
    \mathbf{T}
    \right\|_{\mathrm{F}}
    },
    \label{eq:supp_operator_error}
\end{equation}
where $\|\cdot\|_{\mathrm{F}}$ denotes the Frobenius norm and
$\widetilde{\mathbf{S}}^{\mathrm{sim}}$ is the simulated complex scattering matrix after
normalization by the measured transmission amplitude. The equivalent operator bit
precision is then defined as
\begin{equation}
    b_{\mathrm{op}}
    =
    -\log_{2}
    \left(
    \epsilon_{\mathrm{rms}}
    \right).
    \label{eq:supp_operator_bit}
\end{equation}
This definition follows the intuitive interpretation that one additional bit corresponds to a
factor-of-two reduction in the normalized complex-matrix error. Equivalently, if the matrix
fidelity is reported as
\begin{equation}
    \mathcal{F}_{\mathrm{mat}}
    =
    1-\epsilon_{\mathrm{rms}},
\end{equation}
the operator bit precision can be written as
\begin{equation}
    b_{\mathrm{op}}
    =
    -\log_{2}
    \left(
    1-\mathcal{F}_{\mathrm{mat}}
    \right).
\end{equation}

Using this criterion, the inverse designed non-unitary matrix fdtd simulation test with an average matrix fidelity of
$97.76\%$ corresponds to
\begin{equation}
    b_{\mathrm{op}}
    =
    -\log_{2}(1-0.9776)
    \approx
    5.48~\mathrm{bits}.
\end{equation}
Therefore, although the unitary device library is indexed by a 3-bit MZI control grid, the
actual reconstructed complex operators exhibit an effective matrix precision exceeding
5 bits in this representative validation. This result indicates that the inverse-designed MPU
does not merely implement coarse digital states, but can reproduce the assigned complex
operator with substantially finer analog accuracy within the controlled insertion-loss
budget.
\begin{figure}[ht]
\centering
\includegraphics[width=0.7\textwidth]{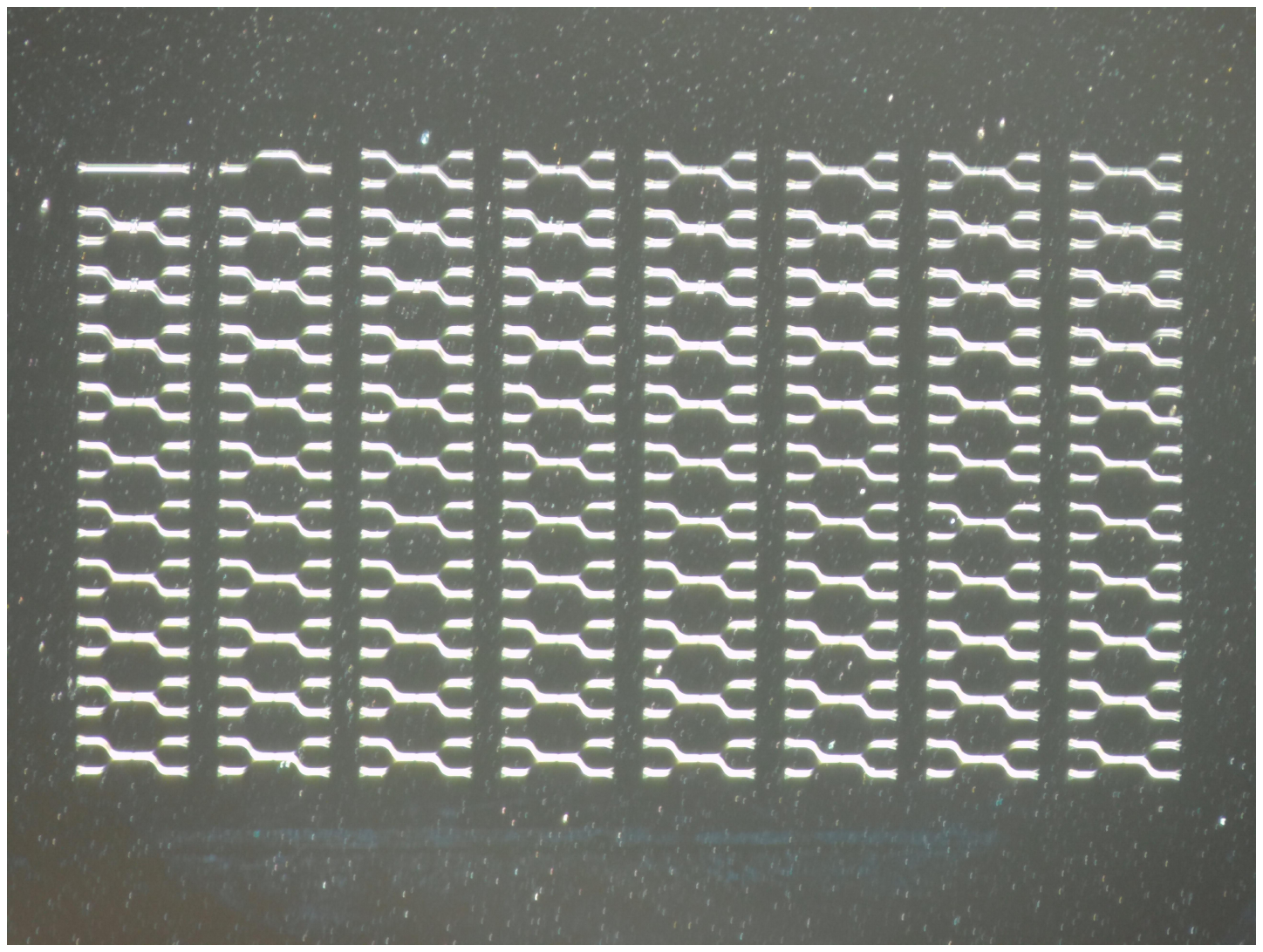}
\caption{\textbf{ Microscopic images of the MPU device library}}
\label{fig:s4}
\end{figure}

\section{CMOS-Compatible Fabrication and Packaging}

\begin{figure}[H]
\centering
\includegraphics[width=0.9\textwidth]{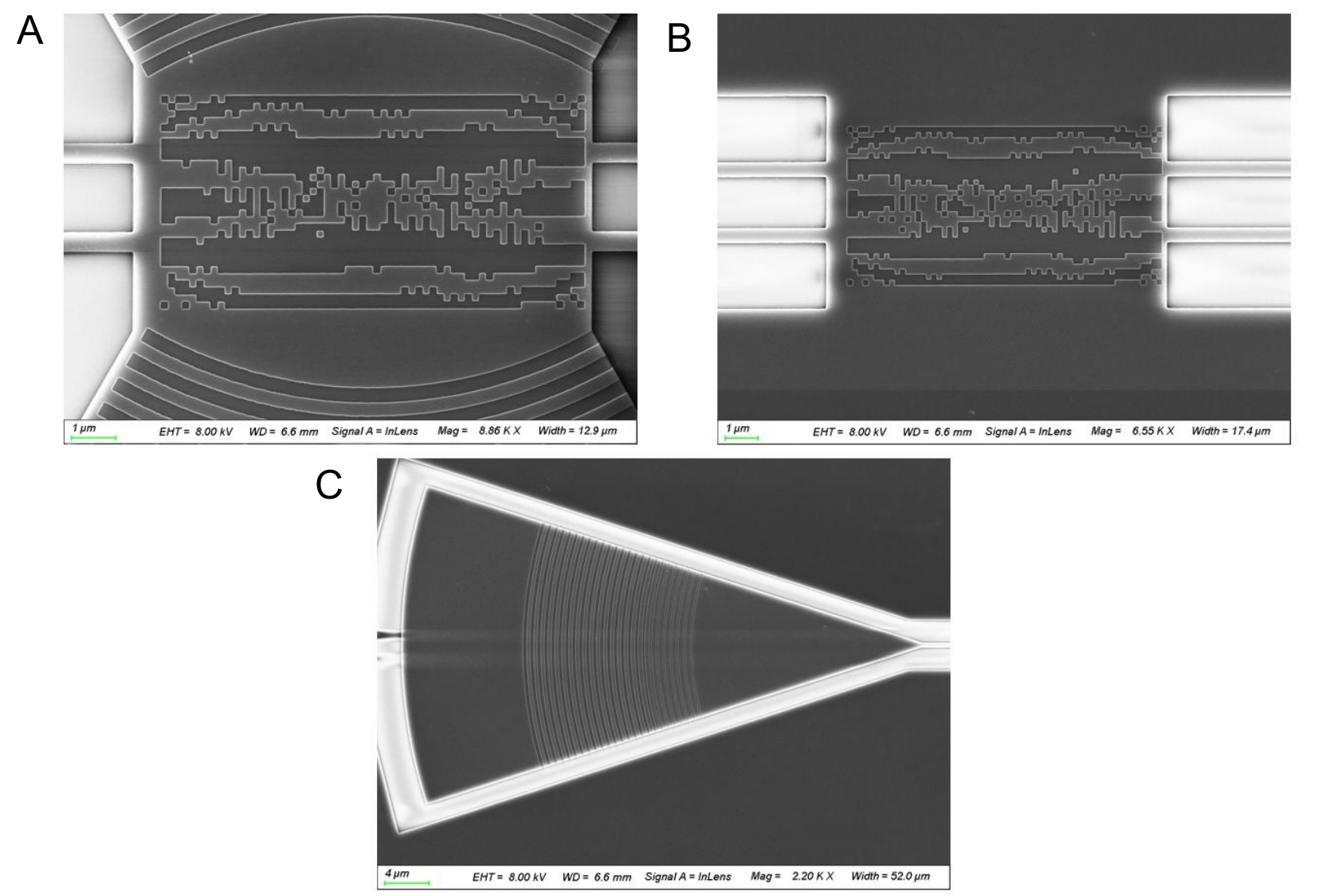}
\caption{
\textbf{SEM images of fabricated inverse-designed optical operators and input grating coupler.}
\textbf{a,b,} Scanning electron microscope images of representative shallow-etched
inverse-designed diffractive operators for $2\times2$ complex matrix transformation.
Two boundary implementations are shown: one using grating-like absorbing boundaries to
suppress lateral scattering and residual feedback, and the other using a slab-waveguide
boundary for a more compact implementation. \textbf{c,} SEM image of the input grating
coupler used for fiber-to-chip optical coupling. The fabricated structures show well-defined
shallow-etched features and smooth interfaces, confirming the CMOS-compatible fabrication
of the inverse-designed photonic operator library.
}
\label{fig:s5}
\end{figure}

\subsection{Fabrication Protocol for MPU divices}

The devices were fabricated on a silicon-on-insulator (SOI) wafer with a
220 nm top silicon layer and a buried oxide (BOX) layer. Before the formal
SOI fabrication, a preliminary 220 nm full-etch process was performed to
verify the electron-beam lithography dose, critical linewidth, and etching
depth of the DOE/MPU patterns. In this test process, the wafer was cleaned,
spin-coated with a 250 nm-thick ZEP electron-beam resist, exposed by
electron-beam lithography (EBL), transferred into the silicon layer by
inductively coupled plasma reactive-ion etching (ICP-RIE) with a 220 nm
etch depth, stripped of resist, and inspected by scanning electron
microscopy (SEM).

The formal SOI fabrication followed a two-step EBL and etching process.
First, the wafer was cleaned and spin-coated with a 250 nm-thick ZEP
positive-tone electron-beam resist. The inverse-designed diffractive
patterns were defined by EBL and transferred into the top silicon layer by
ICP-RIE with a shallow etch depth of 70 nm. This shallow-etched geometry
forms the weakly scattering diffractive nanostructures used for compact
complex-valued optical matrix operations.

After resist removal, a second lithography step was performed to define the
waveguide and device boundary patterns. A new 250 nm-thick ZEP resist layer
was spin-coated, exposed by EBL, and developed. The exposed silicon was then
etched through the full 220 nm top-silicon layer by ICP-RIE to form the
photonic waveguides and isolation trenches. The remaining resist was removed
after etching.

Finally, a 2 $\mu$m-thick silicon dioxide (SiO$_2$) upper cladding layer was
deposited by plasma-enhanced chemical vapor deposition (PECVD), providing
device protection and a stable waveguiding environment. The fabricated
structures were inspected using optical microscopy and SEM to confirm the
pattern fidelity and etch quality.

\subsection{Packaging of Optical Computing Chips}

The photonic integrated circuits were fabricated on a silicon-on-insulator (SOI) platform
using a CMOS-compatible process. The device layer consists of a shallow-etched silicon
photonic layer, which enables low-back-reflection diffractive wavefront manipulation while
remaining compatible with standard passive waveguides, grating couplers, modulators, and
photodetectors. This fabrication strategy allows the inverse-designed meta processing units
(MPUs) to be integrated with conventional silicon photonic building blocks on the same chip.

Supplementary Fig.~S5 shows scanning electron microscope (SEM) images of representative
fabricated devices. The upper panels compare two implementations of the inverse-designed
$2\times2$ diffractive optical operator. In one design, grating-like absorbing boundaries are
introduced around the inverse-designed region to suppress residual lateral scattering and
reduce spurious feedback into the computational window. In the other design, a slab-waveguide
boundary is used, providing a simpler and more compact implementation while maintaining the
dominant forward-propagating field evolution. These two fabricated variants demonstrate that
the MPU layout can be adapted to different boundary treatments according to the requirements
of loss suppression, footprint, and fabrication simplicity. The lower panel shows a fabricated
input grating coupler used for fiber-to-chip optical coupling at the target wavelength. The
clear definition of the grating teeth, waveguide interfaces, and shallow-etched diffractive
features confirms the process compatibility of the inverse-designed subwavelength patterns.

After fabrication, the chip was mounted on a printed circuit board (PCB) carrier for optical
and electrical characterization, as shown in Supplementary Fig.~S6. For experimental
convenience and stable device verification, the measurements reported in this work were
performed using direct-current (DC) electrical interfaces. The DC pads were used to bias the
on-chip Mach--Zehnder modulators (MZMs) and to extract photocurrent signals from the
integrated photodetectors during static and quasi-static system-level characterization. The
photocurrent outputs were routed to external low-noise current-measurement electronics,
while the MZM bias voltages were supplied by programmable DC voltage sources.

Importantly, although the present experiments used a DC-access configuration for simplicity
and reliable calibration, the chip and PCB were designed with high-speed electrical interfaces
compatible with GHz-bandwidth MZM driving. The fabricated high-frequency transmission
lines and RF-compatible contact layout provide a direct path toward high-speed electro-optic
modulation in future system demonstrations. Therefore, the current packaging scheme not
only supports static verification of the inverse-designed photonic operator library, but also
preserves the hardware compatibility required for ultrafast modulation and dynamic photonic
computing.

Supplementary Fig.~S6 further presents microscope images of the packaged chip and the
on-board electrical interconnects. The optical microscope image confirms the successful
fabrication of the multi-device photonic operator library, while the PCB-level image shows
the packaged testing platform with electrical access to the integrated active components.
Together, these results demonstrate the manufacturability, packaging feasibility, and
scalability of the proposed MPU platform toward large-scale integrated photonic computing
systems.

\begin{figure}[ht]
\centering
\includegraphics[width=0.9\textwidth]{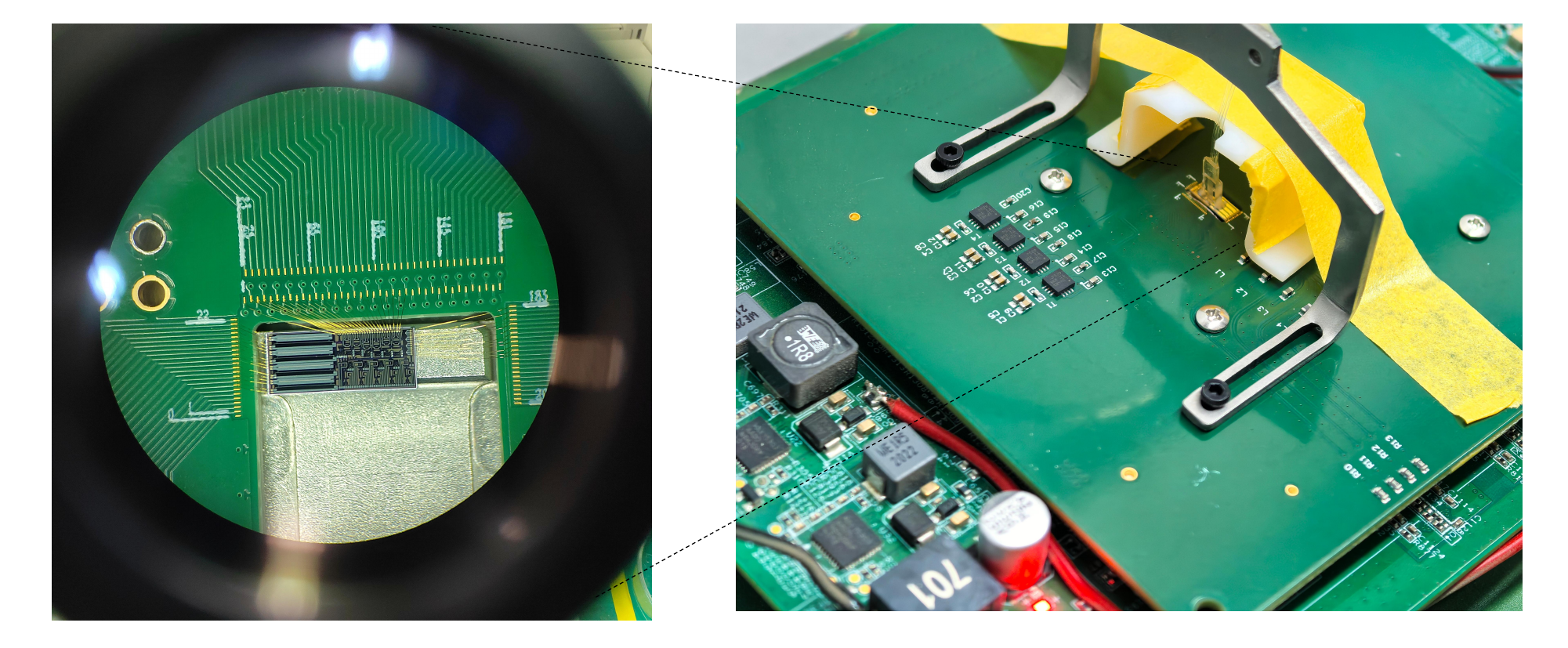}
\caption{
\textbf{Optical microscope image and PCB-level packaging of the fabricated photonic chip.}
}
\label{fig:s6}
\end{figure}

\section{Experimental Characterization Protocols}

\subsection{Photodetector Sensitivity Calibration}

The integrated photodetectors (PDs) were calibrated to establish the quantitative conversion
between the incident optical power and the generated photocurrent. This calibration is
essential for reconstructing optical output intensities, evaluating scattering matrix elements,
and estimating the optical power flow in the photonic computing system.

As shown in Supplementary Fig.~\ref{fig:s7}, we performed linear sensitivity calibration in
two optical-power regimes. For each regime, the input optical power was swept over the
corresponding range, and the output photocurrent was recorded. The measured data were
then fitted using a linear model,
\begin{equation}
    I_{\mathrm{PD}} = R_{\mathrm{sys}} P_{\mathrm{in}} + I_{0},
\end{equation}
where $I_{\mathrm{PD}}$ is the measured photocurrent, $P_{\mathrm{in}}$ is the incident optical
power, $R_{\mathrm{sys}}$ is the system-level responsivity, and $I_0$ is the fitted current
offset.

In the normal-power regime, the calibrated responsivity is
$R_{\mathrm{sys}}=36.67~\mathrm{\mu A/mW}$, with a fitted intercept of
$-5.23~\mathrm{\mu A}$. The corresponding zero-current bias point is approximately
$0.14~\mathrm{mW}$. In the low-power regime, which is relevant to weak optical readout, the
PD maintains a linear response with a responsivity of
$29.82~\mathrm{\mu A/mW}$ and a much smaller intercept of
$-0.16~\mathrm{\mu A}$, corresponding to a near-zero bias threshold of
$0.01~\mathrm{mW}$.

The linear fitting results confirm that the integrated PDs provide stable and reproducible
power-to-current conversion over both conventional and low-power detection ranges. The
calibrated responsivities were used to convert the measured photocurrents into optical
powers in the subsequent device characterization and system-level analysis.

\begin{figure}[H]
\centering
\includegraphics[width=1.0\textwidth]{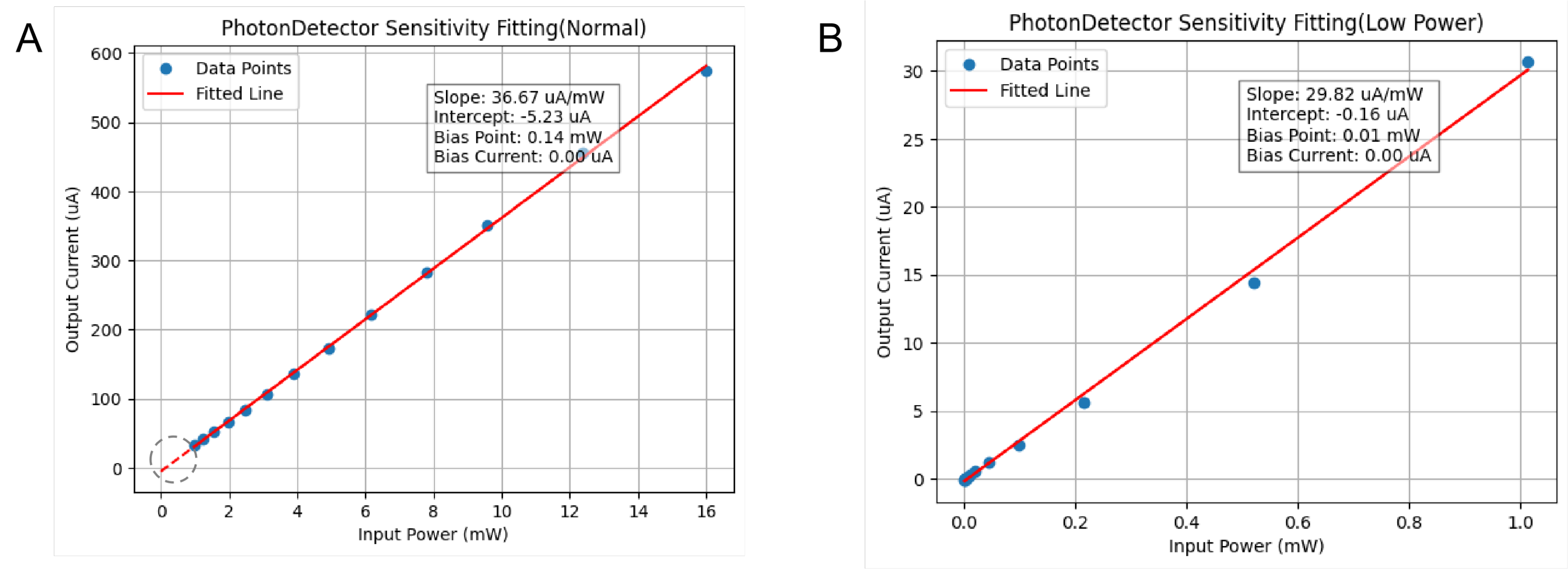}
\caption{
\textbf{Dual-regime sensitivity calibration of the integrated photodetector.}
}
\label{fig:s7}
\end{figure}

\subsection{Scattering Matrix Measurement and wavelength-scanning}
The passive scattering response of the inverse-designed optical operators was characterized
using a wavelength-scanning measurement setup, as illustrated in Supplementary
Fig.~\ref{fig:s8}. A tunable laser source (Ceyear 6317A-H03) was used to provide the input
optical signal. The input polarization was adjusted by a polarization controller to match the
on-chip waveguide mode, and the light was coupled into the selected input port through a
lensed fiber.

For each input excitation, the optical powers from all output ports were measured
simultaneously using a multichannel optical power meter. The laser wavelength was swept
over the target spectral range, and the output responses were recorded at each wavelength.
The measurement was then repeated by sequentially switching the input excitation port. By
combining the output responses from all input excitations, the wavelength-dependent
scattering matrix of the device was reconstructed.

For a two-input--two-output device, the measured transmission matrix at wavelength
$\lambda$ can be written as
\begin{equation}
    \mathbf{S}_{\mathrm{meas}}(\lambda)
    =
    \begin{bmatrix}
    s_{11}(\lambda) & s_{12}(\lambda)\\
    s_{21}(\lambda) & s_{22}(\lambda)
    \end{bmatrix},
\end{equation}
where $s_{ij}(\lambda)$ denotes the optical response measured at output port $i$ when
input port $j$ is excited. The measured powers were corrected using the calibrated PD
responsivity and the reference optical path response. This procedure allows us to quantify
the wavelength-dependent transmission characteristics of the fabricated diffractive optical
operators and to compare the experimental device response with the target scattering
matrix.
\begin{figure}[ht]
\centering
\includegraphics[width=1.0\textwidth]{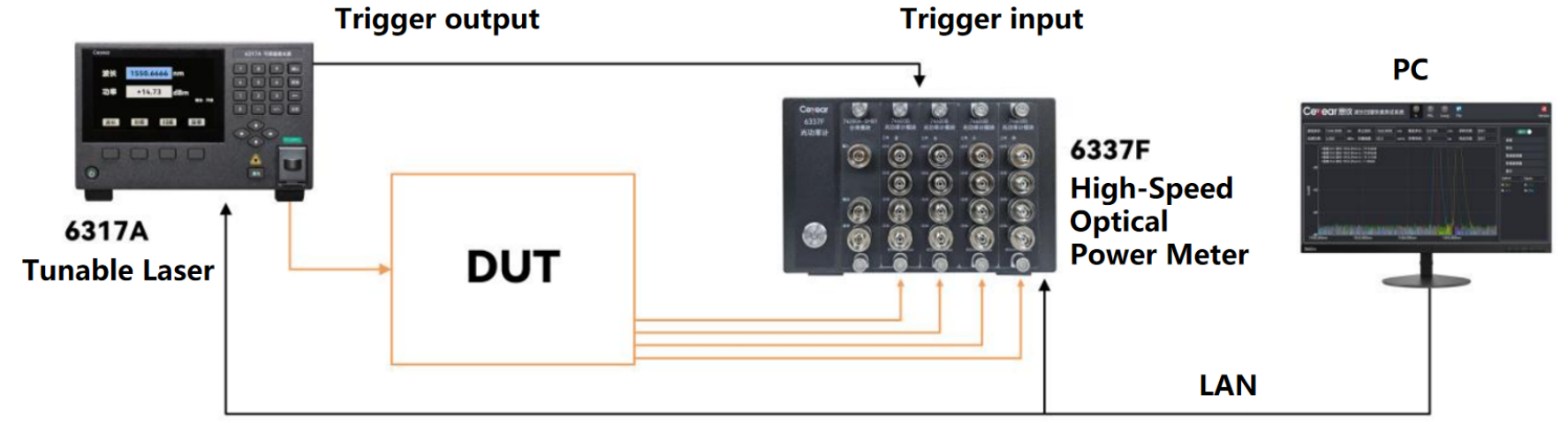}
\caption{
\textbf{Wavelength-scanning setup for scattering matrix characterization.}
A tunable laser source is coupled into the device under test (DUT) through a polarization
controller and a lensed fiber. The optical powers from multiple output ports are recorded
simultaneously by a multichannel optical power meter. By sweeping the laser wavelength
and sequentially exciting different input ports, the full wavelength-dependent scattering
matrix of the inverse-designed optical operator is reconstructed.
}
\label{fig:s8}
\end{figure}

\subsection{Complex-domain matrix reconstruction using on-chip MZI voltage scanning}

\begin{figure}[H]
\centering
\includegraphics[width=0.8\textwidth]{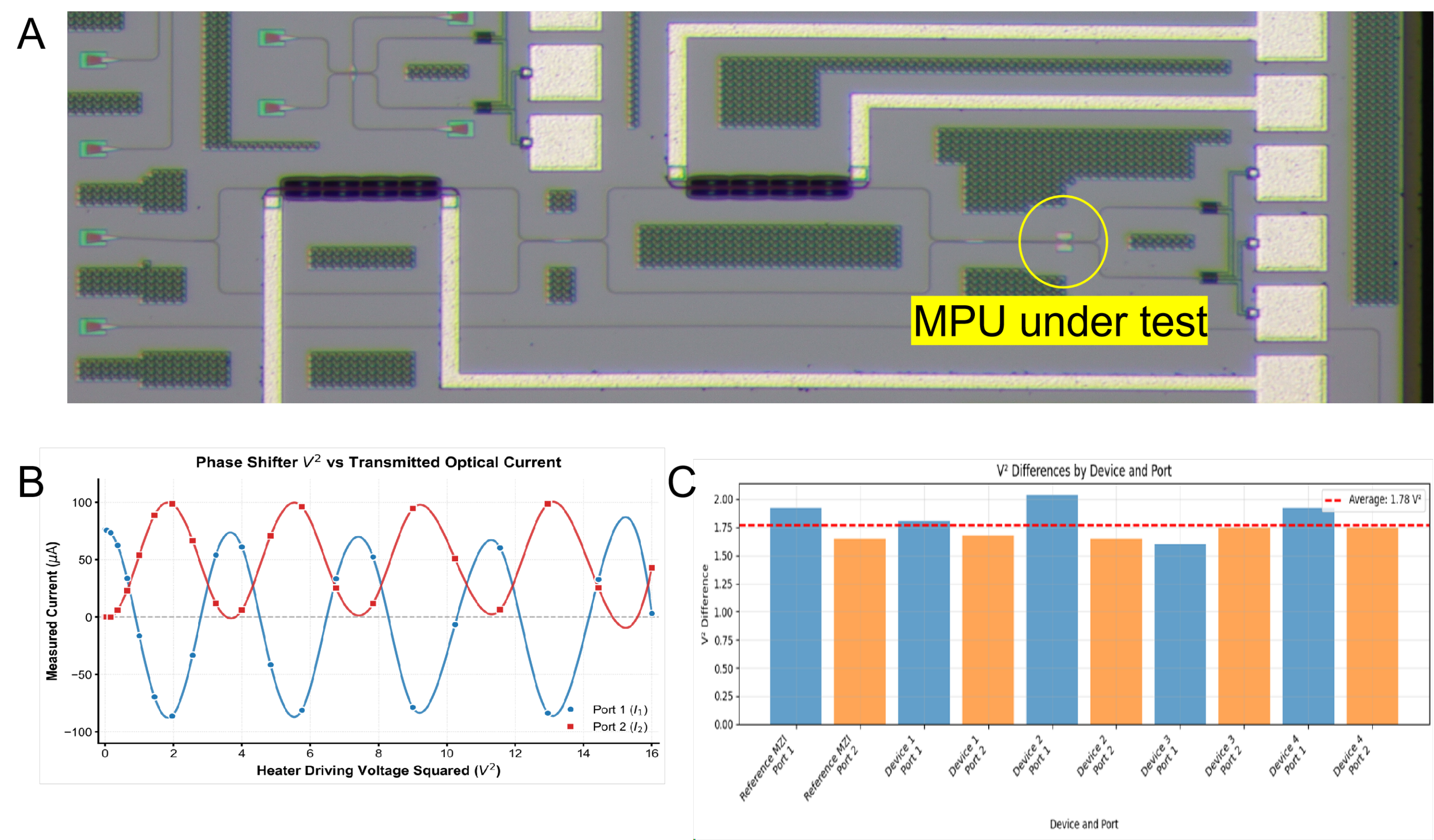}
\caption{
\textbf{On-chip MZI voltage-scan calibration for complex matrix reconstruction.}
\textbf{a,} Optical microscope image of the measurement region, with the MPU under test
highlighted. \textbf{b,} Representative photocurrent responses obtained by sweeping the
voltage applied to the MZI phase shifter. The periodic output currents provide the
voltage-to-phase calibration and the phase reference for complex-domain reconstruction.
\textbf{c,} Extracted half-wave voltages $V_{\pi}$ for multiple devices and ports, showing
the consistency of the fabricated phase-shifting interfaces. By comparing the reference MZI
scan with the cascaded MZI--MPU scan, the amplitude and phase response of each MPU
matrix element can be reconstructed.
}
\label{fig:s9}
\end{figure}

The complex-valued transfer matrix of the fabricated MPU was reconstructed using an
on-chip Mach--Zehnder interferometer (MZI) as a phase-scanning probe. In contrast to a
free-space or external interferometric measurement, the phase reference in our experiment
is provided by the integrated MZI itself. By sweeping the voltage applied to the MZI phase
shifter, the relative phase between the two MZI arms is continuously modulated, leading to
periodic photocurrent responses at the two output ports. These voltage-dependent
photocurrent curves encode the amplitude and phase information of the optical field after
passing through the cascaded MPU.

We first measured a reference MZI without the cascaded MPU. The two output currents of
the reference MZI can be written as
\begin{equation}
    I_{1}^{\mathrm{ref}}(V)
    =
    A_{1}^{\mathrm{ref}}
    +
    B_{1}^{\mathrm{ref}}
    \cos[\Delta\phi(V)+\phi_{1}^{\mathrm{ref}}],
\end{equation}
\begin{equation}
    I_{2}^{\mathrm{ref}}(V)
    =
    A_{2}^{\mathrm{ref}}
    +
    B_{2}^{\mathrm{ref}}
    \cos[\Delta\phi(V)+\phi_{2}^{\mathrm{ref}}],
\end{equation}
where $V$ is the phase-shifter driving voltage, $\Delta\phi(V)$ is the voltage-induced
phase difference, $A_{k}^{\mathrm{ref}}$ is the DC current offset, $B_{k}^{\mathrm{ref}}$ is the
fringe amplitude, and $\phi_{k}^{\mathrm{ref}}$ is the fitted phase offset of output port $k$.
The two output ports show complementary sinusoidal responses, which provide the
calibration of the MZI phase scan and the half-wave voltage $V_{\pi}$.

After inserting the inverse-designed MPU after the MZI, the MZI output field is transformed
by the MPU before photodetection. For a two-output MPU, the voltage-dependent output
currents can be written as
\begin{equation}
    I_{i}^{\mathrm{MPU}}(V)
    =
    A_{i}^{\mathrm{MPU}}
    +
    B_{i}^{\mathrm{MPU}}
    \cos[\Delta\phi(V)+\phi_{i}^{\mathrm{MPU}}],
    \qquad i=1,2 .
\end{equation}
The fitted amplitude $B_{i}^{\mathrm{MPU}}$ reflects the coherent interference strength at
output port $i$, while the fitted phase $\phi_{i}^{\mathrm{MPU}}$ gives the relative phase shift
introduced by the cascaded MPU with respect to the reference MZI scan. Therefore, the
complex response of each output channel can be extracted by comparing the fitted
sinusoidal response of the cascaded MZI--MPU system with that of the reference MZI.

For each input condition, the complex coefficient associated with output port $i$ is obtained
as
\begin{equation}
    \widehat{M}_{ij}
    =
    \gamma_i
    \sqrt{
    \frac{B_{i}^{\mathrm{MPU}}}{B_{i}^{\mathrm{ref}}}
    }
    \exp\left[
    j
    \left(
    \phi_{i}^{\mathrm{MPU}}-\phi_{i}^{\mathrm{ref}}
    \right)
    \right],
    \label{eq:mzi_scan_complex_coeff}
\end{equation}
where $j$ denotes the input port index and $\gamma_i$ is a calibration factor that accounts
for the port-dependent responsivity, coupling imbalance, and static loss of the measurement
path. The square-root relation converts the measured photocurrent modulation depth into
an optical-field amplitude. Repeating the voltage scan for different input excitations yields
all elements of the measured complex matrix,
\begin{equation}
\mathbf{M}_{\mathrm{exp}}
=
\begin{bmatrix}
M_{11} & M_{12}\\
M_{21} & M_{22}
\end{bmatrix},
\qquad
M_{ij}
=
|M_{ij}|e^{j\angle M_{ij}} .
\end{equation}

In practice, each measured photocurrent curve is fitted to the form
\begin{equation}
    I(V)
    =
    A
    +
    B\cos
    \left[
    \frac{\pi}{V_{\pi}}V+\phi
    \right],
    \label{eq:voltage_scan_fit}
\end{equation}
where $A$, $B$, $\phi$, and $V_{\pi}$ are obtained from nonlinear sinusoidal fitting. The
reference MZI scan determines the common voltage-to-phase mapping and removes the
intrinsic phase response of the probe interferometer. The residual phase shift and amplitude
change observed after cascading the MPU are therefore attributed to the complex transfer
function of the MPU. This procedure is illustrated in Fig.~3c of the main text, where the
reference MZI scan and the MZI--MPU scans of several fabricated devices are compared.
The lower polar plots in Fig.~3c show that the reconstructed experimental complex matrix
elements agree with the FDTD-simulated matrix coefficients after global alignment.

To enable a fair comparison between experiment and simulation, we remove the
non-informative global amplitude scale and global phase offset. The measured and simulated
matrices are first represented in the complex domain as
\begin{equation}
\mathbf{M}
=
\begin{bmatrix}
M_{11} & M_{12}\\
M_{21} & M_{22}
\end{bmatrix},
\qquad
M_{ij}=|M_{ij}|e^{j\angle M_{ij}} .
\end{equation}
The experimental matrix is then aligned to the simulated matrix by
\begin{equation}
\mathbf{M}_{\mathrm{exp}}^{\mathrm{aligned}}
=
\alpha \mathbf{M}_{\mathrm{exp}}e^{-j\theta_0},
\end{equation}
where
\begin{equation}
\theta_0
=
\angle
\mathrm{Tr}
\left(
\mathbf{M}_{\mathrm{sim}}^{H}\mathbf{M}_{\mathrm{exp}}
\right),
\end{equation}
and
\begin{equation}
\alpha
=
\frac{
\|\mathbf{M}_{\mathrm{sim}}\|_F
}{
\|\mathbf{M}_{\mathrm{exp}}e^{-j\theta_0}\|_F
}.
\end{equation}
The fitting error is quantified using the normalized Frobenius error,
\begin{equation}
\varepsilon_F
=
\frac{
\left\|
\mathbf{M}_{\mathrm{exp}}^{\mathrm{aligned}}
-
\mathbf{M}_{\mathrm{sim}}
\right\|_F
}{
\left\|
\mathbf{M}_{\mathrm{sim}}
\right\|_F
}.
\end{equation}
This aligned error evaluates the intrinsic agreement of the complex matrix structure,
including the relative amplitude and phase among matrix elements. The absolute
transmission efficiency is reported separately and is not inferred from this globally aligned
matrix error.

\section{Hardware-in-the-loop multi-task classification experiment}

\subsection{Hybrid digital--photonic model and trainable parameters}

To validate the task-level computing capability of the fabricated MPU photonic processor, we
implemented a hardware-in-the-loop classification framework that combines a lightweight
digital front-end, programmable electro-optic input encoding, an on-chip photonic computing
core, and calibrated electronic readout. The complete model contains only 64 trainable
parameters, distributed across both digital and physical domains.

The first part is a digital preprocessing network (PreNet), which maps the raw input feature
vector to a compact four-dimensional representation. For the vowel-classification task used
in this experiment, the input feature vector is a 10-dimensional acoustic feature vector. The
PreNet is implemented as a single fully connected layer followed by a sigmoid activation,
\begin{equation}
    \mathbf{z}
    =
    \sigma
    \left(
    \mathbf{W}_{\mathrm{pre}}\mathbf{x}
    +
    \mathbf{b}_{\mathrm{pre}}
    \right),
    \qquad
    \mathbf{z}\in[0,1]^4 ,
    \label{eq:prenet}
\end{equation}
where $\mathbf{x}$ is the input feature vector, $\mathbf{W}_{\mathrm{pre}}\in\mathbb{R}^{4\times10}$
and $\mathbf{b}_{\mathrm{pre}}\in\mathbb{R}^{4}$ are trainable digital parameters, and
$\sigma(\cdot)$ denotes the sigmoid function. This layer contributes 44 trainable
parameters.

The second part is the electrical-to-optical mapping interface. The normalized PreNet
outputs are converted into physical driving voltages for the on-chip Mach--Zehnder
modulators (MZMs). For the $i$-th input channel, the voltage is given by
\begin{equation}
    V_i
    =
    V_{\mathrm{LO},i}
    +
    \alpha_i
    \left(
    V_{\mathrm{HI},i}
    -
    V_{\mathrm{LO},i}
    \right)
    z_i ,
    \label{eq:mzm_voltage_mapping}
\end{equation}
where $V_{\mathrm{LO},i}$ and $V_{\mathrm{HI},i}$ define the calibrated linear operating
window of the corresponding MZM, and $\alpha_i$ is a trainable channel-dependent scaling
factor. The four $\alpha_i$ parameters allow the model to adapt the digital feature
amplitudes to the actual modulation depth and nonideal transfer characteristics of the
hardware.

The third part is the photonic computing core. The encoded optical signals propagate
through the MPU-based photonic network and are coherently transformed by the physical
interference and diffraction processes. In the hardware-in-the-loop experiment, the
configurable optical core is parameterized by 12 trainable phase settings,
\begin{equation}
    \boldsymbol{\theta}
    =
    [\theta_1,\theta_2,\ldots,\theta_{12}],
\end{equation}
which tune the on-chip interferometric phase distribution and determine the effective
optical transformation implemented by the processor.

Finally, the optical outputs are converted into electrical currents by the integrated
photodetectors. To compensate channel-dependent responsivity variations, coupling
imbalance, dark-current offsets, and static loss differences, a trainable detector calibration
factor is assigned to each output channel,
\begin{equation}
    \mathbf{I}_{\mathrm{cal}}
    =
    \boldsymbol{\beta}
    \odot
    \mathbf{I}_{\mathrm{raw}},
    \label{eq:pd_calibration}
\end{equation}
where $\mathbf{I}_{\mathrm{raw}}$ is the measured photocurrent vector,
$\boldsymbol{\beta}\in\mathbb{R}^{4}$ is the trainable readout scaling vector, and
$\odot$ denotes element-wise multiplication. The calibrated outputs are then normalized and
processed by the digital backend to obtain the classification result.

The complete trainable parameter vector is therefore
\begin{equation}
    \boldsymbol{\Theta}
    =
    \{
    \mathbf{W}_{\mathrm{pre}},
    \mathbf{b}_{\mathrm{pre}},
    \boldsymbol{\alpha},
    \boldsymbol{\theta},
    \boldsymbol{\beta}
    \},
    \label{eq:hil_trainable_parameters}
\end{equation}
including 44 PreNet parameters, 4 input-scaling parameters, 12 physical phase parameters,
and 4 readout-calibration parameters.

\begin{figure}[ht]
\centering
\includegraphics[width=1.0\textwidth]{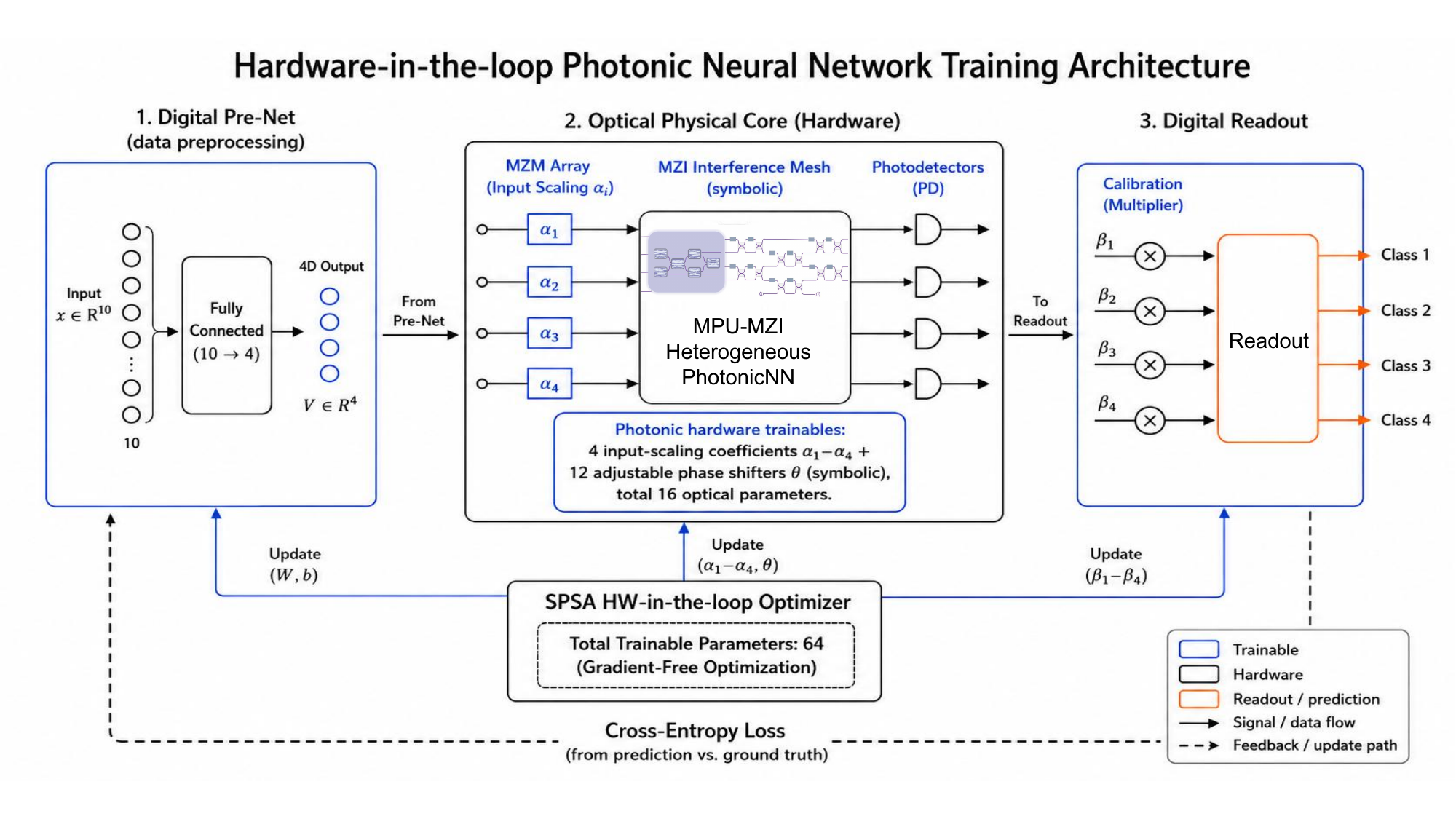}
\caption{
\textbf{Hardware-in-the-loop training architecture of the hybrid photonic neural network.}
The architecture combines a digital PreNet, an MPU--MZI photonic physical core, and a
digital readout module. The PreNet compresses the 10-dimensional input feature vector into
a 4-dimensional optical input representation. The optical core performs physical
computation using an MZM input-scaling array, an MZI interference mesh integrated with
the MPU photonic operator, and on-chip photodetectors. The readout module applies
channel-wise calibration factors to compensate detector and path imbalance before
generating four class outputs. The full model contains 64 trainable parameters: 44 digital
PreNet parameters $(W,b)$, 4 input-scaling coefficients $\alpha$, 12 physical phase-shifter
parameters $\theta$, and 4 readout calibration factors $\beta$. The cross-entropy loss is
computed from the measured hardware output, and an SPSA-based gradient-free optimizer
updates all trainable parameters in the hardware-in-the-loop training process. Solid arrows
denote signal/data flow and dashed arrows denote feedback/update paths.
}
\label{fig:s10}
\end{figure}

\subsection{Mach--Zehnder modulator calibration}

Before hardware-in-the-loop training, the MZM transfer functions were calibrated to identify
stable and approximately monotonic voltage windows for analog optical modulation. This
step is necessary because the electrical voltages applied to the MZMs serve as the physical
interface between the digital PreNet output and the optical input amplitudes of the
photonic processor.

As shown in Supplementary Fig.~\ref{fig:s11}a, repeated voltage sweeps from 0 to 3 V were
applied to the MZM driving electrodes, and the corresponding optical transmission was
recorded. The measured response exhibits a periodic modulation curve, consistent with the
interferometric transfer function of an MZM. The repeated cycles verify stable modulation
behavior and reproducible extinction.

From the measured voltage-transmission curve, we extracted a quasi-linear operating
region between the transmission peak and null. This region was selected to provide a
single-valued and approximately linear mapping from voltage to optical intensity, avoiding
phase wrapping or non-monotonic modulation that would otherwise confuse the
hardware-in-the-loop update. As shown in Supplementary Fig.~\ref{fig:s11}b, linear
regression of the selected operating region gives a coefficient of determination
$R^2=0.9816$, confirming that the MZM can be used as a predictable analog optical
encoder within this voltage range. The extracted lower and upper voltage bounds,
$V_{\mathrm{LO}}$ and $V_{\mathrm{HI}}$, were used in Eq.~\eqref{eq:mzm_voltage_mapping}
to constrain all subsequent driving voltages during training and inference.

\begin{figure}[ht]
\centering
\includegraphics[width=0.8\textwidth]{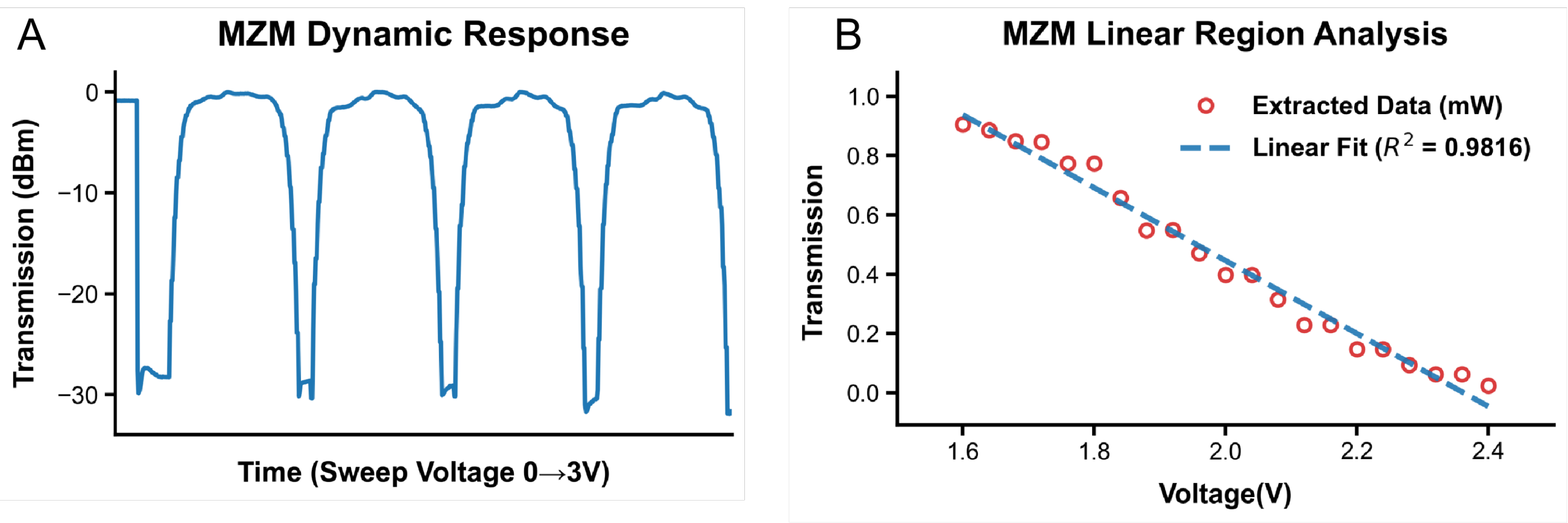}
\caption{
\textbf{Calibration of the on-chip Mach--Zehnder modulator.}
\textbf{a,} Dynamic optical transmission response of the MZM under repeated 0--3 V
voltage sweeps. The periodic curve verifies stable electro-optic modulation and provides the
full voltage-transmission characteristic. \textbf{b,} Linear-region extraction and fitting of the
normalized transmission as a function of driving voltage. The selected operating region
exhibits an approximately linear response with $R^2=0.9816$, defining the voltage window
used for analog optical input encoding.
}
\label{fig:s11}
\end{figure}

\subsection{Hardware calibration and signal normalization}

Prior to closed-loop training, the optical and electrical channels were calibrated to reduce
static hardware imbalance. The calibration includes three steps. First, the MZM voltage
sweeps described above determine the monotonic operating intervals and the approximate
half-wave modulation scale. Second, the static transmission levels of different optical paths
and detector channels are measured to estimate channel-dependent loss and responsivity
variations. These variations are compensated by the trainable readout scaling factors
$\boldsymbol{\beta}$ and by software-side normalization. Third, the measured photocurrent
statistics are used to stabilize the numerical scale entering the digital classification
backend.

Specifically, for each output channel, a reference set of photocurrent measurements is used
to estimate the mean and standard deviation,
\begin{equation}
    \mu_k = \mathbb{E}[I_k],
    \qquad
    \sigma_k =
    \sqrt{\mathbb{E}[(I_k-\mu_k)^2]} .
\end{equation}
The normalized readout is then computed as
\begin{equation}
    \widehat{I}_k
    =
    \frac{
    I_{\mathrm{cal},k}-\mu_k
    }{
    \sigma_k+\epsilon
    },
    \label{eq:current_normalization}
\end{equation}
where $\epsilon$ is a small numerical constant. This normalization prevents a single
high-power or high-responsivity channel from dominating the classifier and improves the
stability of hardware-in-the-loop optimization.
\subsection{Hardware-in-the-loop setup for multi-task classification}

The hardware-in-the-loop experimental platform is shown in Supplementary
Fig.~\ref{fig:s12}. The system consists of a continuous-wave laser source, the packaged
MPU photonic chip, a multi-channel programmable voltage source, a picoampere-level
photocurrent acquisition unit, and a host computer for control and optimization.

A continuous-wave laser at the telecom wavelength of 1550 nm provides the optical carrier
for the chip. The laser light is coupled into the on-chip photonic circuit and modulated by
the calibrated MZMs. The MZM driving voltages are generated by a multi-channel
programmable voltage source according to the current trainable parameter vector
$\boldsymbol{\Theta}$. The modulated optical signals then propagate through the MPU
photonic computing network, where the physical optical transformation is performed by the
fabricated diffractive and interferometric structures.

The output optical intensities are detected by integrated photodetectors and converted into
photocurrents. The photocurrents are measured by a picoammeter and transferred to the host
computer. The host computer performs signal normalization, computes the classification
loss, estimates the update direction of the trainable parameters, and writes the updated
voltages back to the programmable source. This forms a closed loop between the physical
photonic chip and the digital optimization algorithm.

Unlike an offline simulation-only training pipeline, this workflow directly incorporates the
measured response of the fabricated chip, including fabrication-induced phase errors,
modulator nonlinearity, channel-dependent insertion loss, detector responsivity imbalance,
thermal drift, and coupling fluctuations. The optimized parameters are therefore adapted to
the actual hardware response of the chip.

\begin{figure}[ht]
\centering
\includegraphics[width=1.0\textwidth]{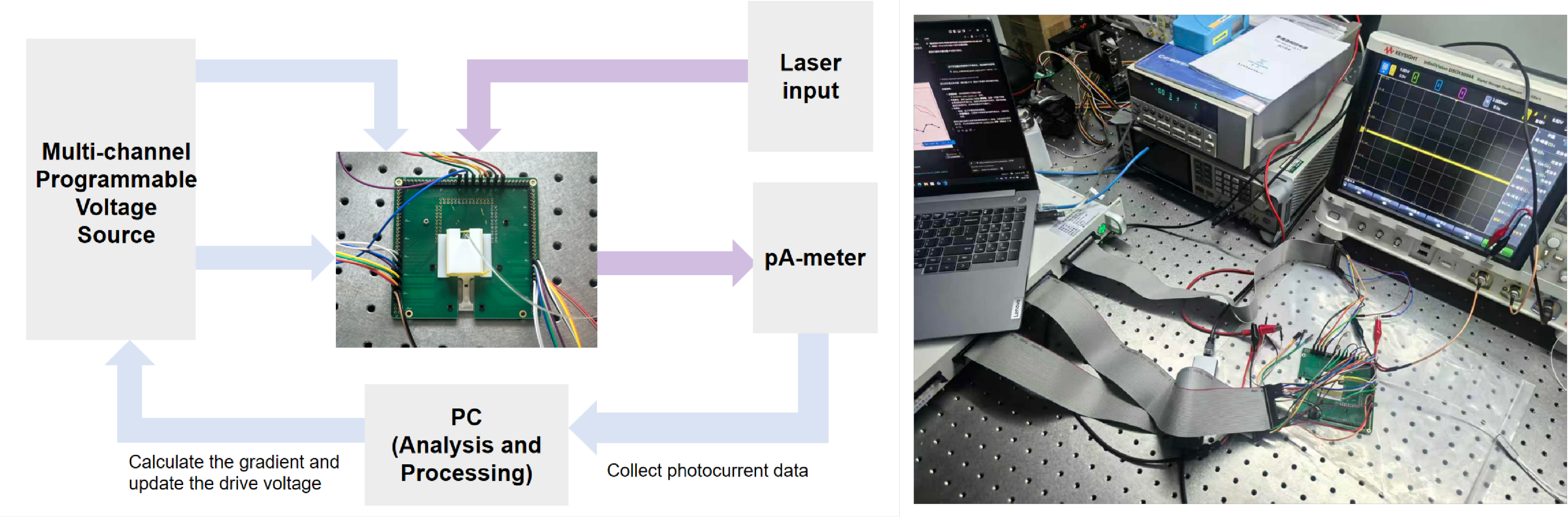}
\caption{
\textbf{Hardware-in-the-loop experimental platform for multi-task photonic classification.}
\textbf{Left,} Schematic of the closed-loop workflow. A 1550-nm laser provides the optical
carrier, the multi-channel programmable voltage source drives the on-chip MZMs and phase
settings, the MPU photonic chip performs optical computation, and the output photocurrents
are measured by a picoammeter. The host PC processes the measured signals, evaluates the
classification loss, and updates the hardware parameters for the next iteration. \textbf{Right,}
Photograph of the physical experimental setup, including the packaged photonic chip,
electrical driving units, photocurrent acquisition equipment, and host-control computer.
}
\label{fig:s12}
\end{figure}

\subsection{SPSA-based hardware-in-the-loop optimization}

Because the fabricated photonic chip is a complex physical system with unknown
fabrication errors and nonideal transfer functions, its exact analytical gradient is not
available. A conventional finite-difference method would require two measurements per
trainable parameter at each optimization step. For the 64-parameter model used here, this
would require at least 128 hardware evaluations per step, which is inefficient for
instrument-limited experiments. We therefore use simultaneous perturbation stochastic
approximation (SPSA) to estimate the gradient with only two hardware evaluations per
iteration, independent of the number of parameters.

At iteration $t$, all trainable digital and physical parameters are concatenated into
$\boldsymbol{\Theta}_t$. A random perturbation vector
$\boldsymbol{\Delta}_t$ is sampled with independent Bernoulli entries,
\begin{equation}
    \Delta_{t,i}\in\{-1,+1\}.
\end{equation}
Two perturbed parameter settings are then applied to the hardware,
\begin{equation}
    \boldsymbol{\Theta}_{t}^{+}
    =
    \boldsymbol{\Theta}_{t}
    +
    c_t\boldsymbol{\Delta}_t,
    \qquad
    \boldsymbol{\Theta}_{t}^{-}
    =
    \boldsymbol{\Theta}_{t}
    -
    c_t\boldsymbol{\Delta}_t,
\end{equation}
where $c_t$ is the perturbation magnitude. The physical chip is evaluated under both
settings, giving two measured losses,
\begin{equation}
    L_t^{+}
    =
    L(\boldsymbol{\Theta}_{t}^{+}),
    \qquad
    L_t^{-}
    =
    L(\boldsymbol{\Theta}_{t}^{-}) .
\end{equation}
The SPSA gradient estimate is computed element-wise as
\begin{equation}
    \widehat{\mathbf{g}}_t
    =
    \frac{
    L_t^{+}-L_t^{-}
    }{
    2c_t
    }
    \boldsymbol{\Delta}_t^{-1},
    \label{eq:spsa_gradient}
\end{equation}
where $\boldsymbol{\Delta}_t^{-1}$ denotes element-wise reciprocal, which is equal to
$\boldsymbol{\Delta}_t$ for Bernoulli $\pm1$ perturbations. The parameter vector is then
updated by
\begin{equation}
    \boldsymbol{\Theta}_{t+1}
    =
    \boldsymbol{\Theta}_{t}
    -
    a_t \widehat{\mathbf{g}}_t ,
    \label{eq:spsa_update}
\end{equation}
where $a_t$ is the learning rate. After the update, the MZM driving voltages are clipped to
their calibrated linear operating windows, and the detector scaling factors are constrained
to remain positive.

The full hardware-in-the-loop procedure is summarized as follows:
\begin{enumerate}
    \item Calibrate the MZM voltage-transmission curves and determine the linear operating
    windows $[V_{\mathrm{LO}},V_{\mathrm{HI}}]$.
    \item Measure baseline photocurrent statistics and initialize the normalization and
    readout-calibration parameters.
    \item Initialize the 64 trainable parameters
    $\boldsymbol{\Theta}=
    \{\mathbf{W}_{\mathrm{pre}},\mathbf{b}_{\mathrm{pre}},
    \boldsymbol{\alpha},\boldsymbol{\theta},\boldsymbol{\beta}\}$.
    \item Generate the positive and negative SPSA perturbations
    $\boldsymbol{\Theta}^{+}$ and $\boldsymbol{\Theta}^{-}$.
    \item Apply the corresponding voltages to the chip, measure the photocurrent outputs,
    and compute the two classification losses $L^{+}$ and $L^{-}$ on the host PC.
    \item Estimate the stochastic gradient using Eq.~\eqref{eq:spsa_gradient} and update
    the parameters using Eq.~\eqref{eq:spsa_update}.
    \item Repeat the measurement--update loop until the validation loss and classification
    accuracy converge.
\end{enumerate}

This SPSA-based strategy enables practical optimization of the hybrid digital--photonic
model using only two physical measurements per iteration. As a result, the training process
can tolerate hardware nonidealities without requiring an explicit differentiable model of the
entire optical system.

\section{Large-scale simulation framework for shared-MPU photonic networks}

\subsection{EMNIST multi-task setting and optical network configuration}

To evaluate the scalability of the proposed MPU-based photonic computing framework, we
performed a large-scale numerical experiment on multi-task EMNIST classification. The
experiment includes three related visual recognition tasks: digit classification with
10 classes, uppercase-letter classification with 26 classes, and lowercase-letter
classification with 26 classes. Each $28\times28$ input image is flattened and projected
onto a 64-dimensional optical input vector using a fixed PCA preprocessing layer. The
optical computing core is modeled as a coherent $64\times64$ Clements mesh, followed by
square-law intensity detection, shared normalization and nonlinear activation layers, and
task-specific electronic classification heads:
\begin{equation}
\begin{aligned}
\mathrm{EMNIST}\;28\times28
&\rightarrow \mathrm{PCA}(784\rightarrow64)
\rightarrow \mathrm{Clements}_{64\times64}
\rightarrow |E|^2 \\
&\rightarrow \mathrm{LayerNorm}
\rightarrow \mathrm{GELU}
\rightarrow \mathrm{task\text{-}specific\;head}.
\end{aligned}
\end{equation}
The task-specific heads isolate the output label spaces of the three tasks, so that the
comparison mainly reflects the optical-core sharing strategy rather than competition among
different classifier outputs.

The Clements mesh consists of alternating even and odd $2\times2$ optical coupling
stages. In our parameterization, the phase tensors are
\begin{equation}
\theta_{\mathrm{even}},\phi_{\mathrm{even}}\in\mathbb{R}^{32\times32},
\qquad
\theta_{\mathrm{odd}},\phi_{\mathrm{odd}}\in\mathbb{R}^{32\times31}.
\end{equation}
Therefore, the total number of candidate replaceable local optical units is
\begin{equation}
N_{\mathrm{unit}}
=
32\times32+32\times31
=
2016 .
\end{equation}
Each candidate unit corresponds to one complete $2\times2$ optical transformation. Thus,
a replacement decision applies jointly to the corresponding $\theta$ and $\phi$ parameters,
which is consistent with the physical interpretation that a passive MPU replaces a complete
local $2\times2$ operator rather than an isolated phase shifter.

\begin{figure}[ht]
\centering
\includegraphics[width=1.0\textwidth]{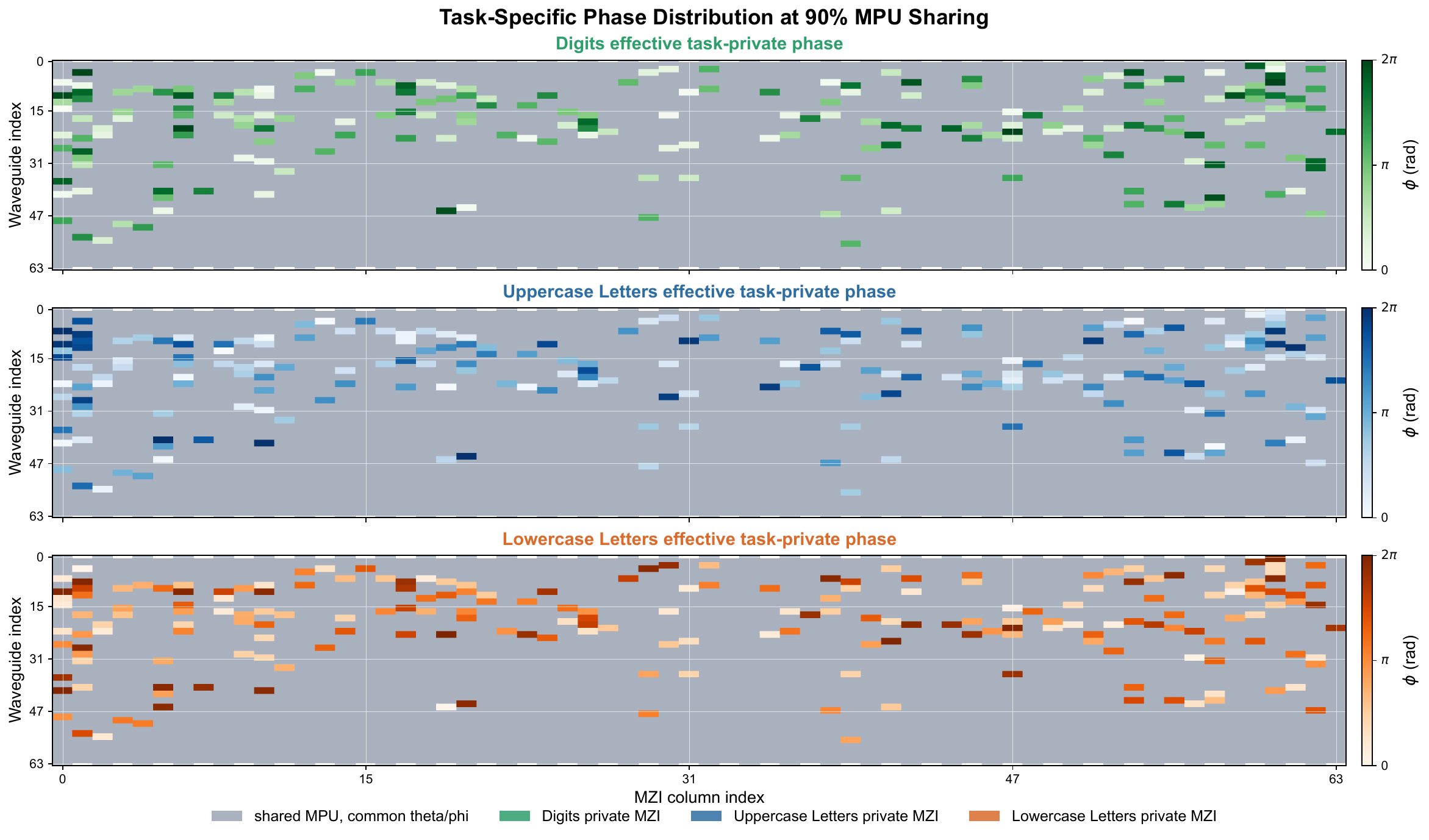}
\caption{
\textbf{Task-specific phase distribution at 90\% shared-MPU replacement.}
Effective task-private phase maps of the $64\times64$ Clements mesh for digit,uppercase-letter, and lowercase-letter classification. Gray regions denote shared MPU neurons, where all tasks use the same common $(\theta,\phi)$ setting. Colored regions denote the remaining task-private MZI nodes, with the color intensity representing the effective phase value. At 90\% MPU sharing, only a sparse subset of optical units remains
task-specific, and these units are non-uniformly distributed across the mesh, indicating that the neuron-level replacement algorithm selectively preserves task-sensitive phase degrees of freedom.
}\label{fig:s13}
\end{figure}

\subsection{Shared-MPU replacement semantics}

Each optical unit $i$ contains a common phase setting $(\theta_i,\phi_i)$ and
task-specific phase offsets $(\Delta\theta_{t,i},\Delta\phi_{t,i})$ for task $t$. The
effective task-dependent setting is
\begin{equation}
\theta_{t,i}^{\mathrm{eff}}
=
\theta_i+\Delta\theta_{t,i},
\qquad
\phi_{t,i}^{\mathrm{eff}}
=
\phi_i+\Delta\phi_{t,i}.
\end{equation}
A binary replacement mask $M_i$ determines whether the unit is implemented as a shared
passive MPU or retained as a task-private MZI,
\begin{equation}
M_i =
\begin{cases}
1, & \mathrm{shared\ MPU},\\
0, & \mathrm{task\text{-}private\ MZI}.
\end{cases}
\end{equation}
When $M_i=1$, the task-specific offsets are disabled and all tasks use the same local
$2\times2$ transformation. When $M_i=0$, the offsets remain enabled and each task can
independently tune the local optical response. The MPU sharing ratio therefore continuously
interpolates between a fully task-private programmable MZI mesh and a fully shared
passive-MPU mesh.

The training procedure contains two stages. First, a warmup model is trained with
$0\%$ MPU sharing, where all optical units remain task-private:
\begin{equation}
M_i=0,\qquad i=1,\ldots,N_{\mathrm{unit}}.
\end{equation}
This stage provides a high-performance task-private reference and reveals which optical
units are naturally consistent or inconsistent across tasks. Second, for each target sharing
ratio $r$, a binary replacement mask is imposed and the model is fine-tuned from the warmup
checkpoint. Shared MPU nodes have their task-specific offsets disabled, whereas
task-private MZI nodes retain task-specific offsets. The replacement mask remains fixed
during fine-tuning. In the final evaluation, the sharing ratio is swept from $0\%$ to $100\%$,
allowing us to compare different replacement strategies under the same passive-MPU budget.

\subsection{Layer-level and neuron-level replacement strategies}

As a coarse baseline, we use a stage-level replacement strategy. For a target sharing ratio
$r$, Clements stages are selected sequentially from the input side until approximately
$rN_{\mathrm{unit}}$ optical units are replaced by shared MPUs. This rule reflects a simple
physical prior: early optical stages are treated as a common feature-mixing backbone,
whereas later stages are reserved for task-specific adaptation.

However, such coarse stage-level replacement does not distinguish whether individual
optical units are truly shared across tasks. We therefore introduce a neuron-level
replacement strategy that refines the stage-level prior using a Fisher-style replacement
damage criterion. The goal is to identify optical units whose task-specific offsets can be
removed with minimal estimated loss increase. A good shared-MPU candidate should be
useful for the learned optical representation, consistent across tasks, and inexpensive to
collapse into a common $2\times2$ transformation.

For a candidate unit $i$, replacing the task-specific MZI settings by a shared MPU requires
collapsing the effective task-specific phases into one common setting. Because optical
phases are periodic, the common phases are computed by circular averaging:
\begin{equation}
\bar{\theta}_i
=
\mathrm{atan2}
\left(
\frac{1}{T}\sum_{t=1}^{T}\sin\theta_{t,i}^{\mathrm{eff}},
\frac{1}{T}\sum_{t=1}^{T}\cos\theta_{t,i}^{\mathrm{eff}}
\right),
\end{equation}
\begin{equation}
\bar{\phi}_i
=
\mathrm{atan2}
\left(
\frac{1}{T}\sum_{t=1}^{T}\sin\phi_{t,i}^{\mathrm{eff}},
\frac{1}{T}\sum_{t=1}^{T}\cos\phi_{t,i}^{\mathrm{eff}}
\right),
\end{equation}
where $T=3$ for the digit, uppercase-letter, and lowercase-letter tasks.

The cross-task phase agreement is measured by the circular concentration
\begin{equation}
R_{\theta,i}
=
\left|
\frac{1}{T}
\sum_{t=1}^{T}
e^{j\theta_{t,i}^{\mathrm{eff}}}
\right|,
\qquad
R_{\phi,i}
=
\left|
\frac{1}{T}
\sum_{t=1}^{T}
e^{j\phi_{t,i}^{\mathrm{eff}}}
\right|.
\end{equation}
The consensus weight is then defined as
\begin{equation}
w_i^{\mathrm{cons}}
=
0.10+
0.90
\left[
\frac{1}{2}
\left(
R_{\theta,i}+R_{\phi,i}
\right)
\right]^2 .
\end{equation}
A large $w_i^{\mathrm{cons}}$ indicates that the effective phase settings of the three tasks
are already close to each other, making the unit a natural candidate for shared-MPU
replacement.

The importance of each optical unit is estimated using a diagonal Fisher proxy from the
warmup model,
\begin{equation}
F_{t,i}
=
\left(
\frac{\partial \mathcal{L}_t}{\partial \theta_i}
\right)^2
+
\left(
\frac{\partial \mathcal{L}_t}{\partial \phi_i}
\right)^2,
\end{equation}
where $\mathcal{L}_t$ is the cross-entropy loss of task $t$. The phase mismatch introduced
by collapsing the task-specific settings into the shared circular mean is
\begin{equation}
d_{t,i}
=
\frac{1}{2}
\left[
\mathrm{wrap}
\left(
\theta_{t,i}^{\mathrm{eff}}-\bar{\theta}_i
\right)^2
+
\mathrm{wrap}
\left(
\phi_{t,i}^{\mathrm{eff}}-\bar{\phi}_i
\right)^2
\right],
\end{equation}
where
\begin{equation}
\mathrm{wrap}(x)=\mathrm{atan2}(\sin x,\cos x).
\end{equation}
The estimated replacement damage is
\begin{equation}
D_i
=
\frac{1}{T}
\sum_{t=1}^{T}
F_{t,i}d_{t,i}.
\end{equation}
After mesh-wide normalization of the Fisher utility and replacement damage as
$\widetilde{U}_i$ and $\widetilde{D}_i$, the final commonness score is defined as
\begin{equation}
S_i
=
w_i^{\mathrm{cons}}
\frac{
0.25+\widetilde{U}_i
}{
0.25+\widetilde{D}_i
}.
\end{equation}
High-scoring nodes are useful, cross-task consistent, and low-damage shared-MPU
candidates. Low-scoring nodes are more suitable to remain task-private MZIs.

For a target sharing ratio $r$, the neuron-level mask is obtained by score-based refinement
of the stage-level prior. Low-score units inside the initially shared front-stage region are
swapped back to task-private MZIs, whereas high-score private units outside the region are
replaced by shared MPUs. The total number of shared units is preserved:
\begin{equation}
\sum_{i=1}^{N_{\mathrm{unit}}}M_i^{\mathrm{node}}
=
\sum_{i=1}^{N_{\mathrm{unit}}}M_i^{\mathrm{stage}}
\approx
rN_{\mathrm{unit}} .
\end{equation}
Thus, the stage-level and neuron-level strategies use the same passive-MPU budget and
the same number of remaining task-private MZIs. Any performance difference is therefore
attributed to the placement of shared MPUs rather than to the total sharing ratio.

\subsection{Visualization of task-specific phases and optical energy flow}

To interpret the learned replacement pattern, we visualize the $90\%$ shared-MPU case in
terms of the remaining task-private phase distribution and the optical energy flow through
the mesh. Supplementary Fig.~S13 shows the effective task-private phase maps for the
digit, uppercase-letter, and lowercase-letter tasks. Gray regions correspond to shared MPU
nodes, where task-specific phase offsets are disabled and the common $(\theta,\phi)$
setting is used by all tasks. Colored regions indicate the remaining task-private MZI nodes,
with the color intensity representing the effective phase value. The three maps show that
only a sparse subset of nodes remains task-specific at the $90\%$ sharing ratio, and these
nodes are distributed non-uniformly across the mesh. This confirms that the neuron-level
algorithm does not simply preserve an entire late-stage block, but selectively retains
task-sensitive optical units where task-dependent phase adaptation is most beneficial.

Supplementary Fig.~S14 further visualizes the normalized optical energy flow through the
same $90\%$ shared-MPU mesh. The gray dots denote shared MPU nodes, the purple dots
denote task-private MZI nodes, and the purple trajectories indicate dominant optical energy
paths. The energy flow is not uniformly distributed over the full Clements mesh; instead,
it concentrates along several structured routes. The remaining task-private MZIs tend to
appear near these high-energy paths, where local phase perturbations have a stronger effect
on the final output. This provides an intuitive physical explanation for the Fisher-style
criterion: replacing a low-energy or cross-task-consistent node by a shared MPU is unlikely
to damage task performance, whereas preserving task-private tunability near important
energy-carrying paths helps maintain classification accuracy even at high MPU sharing
ratios.

Together, these two visualizations show that large-scale shared-MPU photonic networks can
be made highly sparse in task-specific optical parameters without random or uniform
pruning. Instead, the neuron-level replacement rule preserves task-private MZI flexibility
only where it is most useful, while replacing the majority of the mesh by compact shared
MPU primitives. This supports the scalability of the proposed architecture for large
multi-input and multi-output photonic neural networks.

\begin{figure}[ht]
\centering
\includegraphics[width=1.0\textwidth]{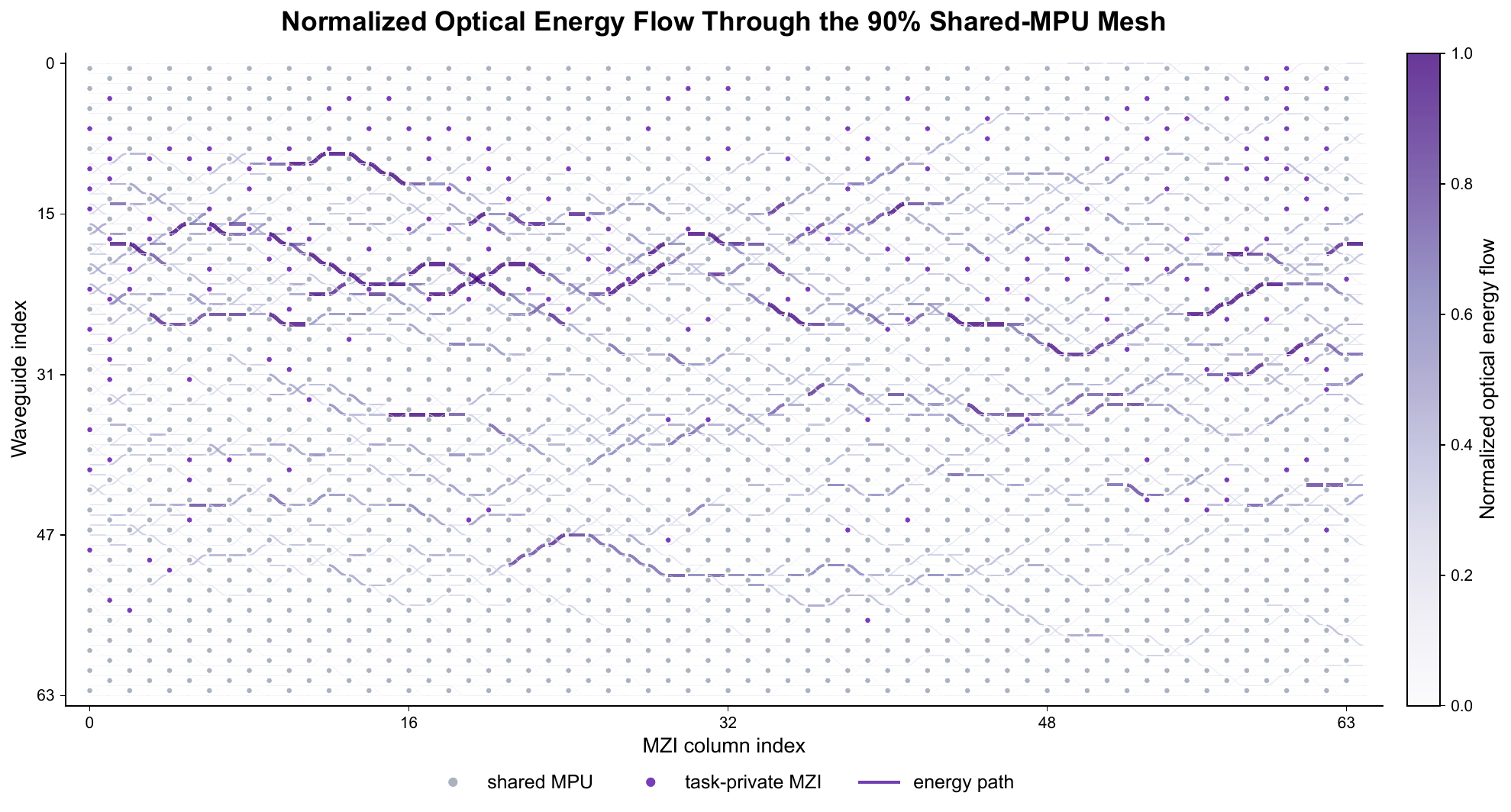}
\caption{
\textbf{Normalized optical energy flow through the 90\% shared-MPU mesh.}
Visualization of the dominant optical energy paths in the high-sharing photonic mesh.
Gray dots represent shared MPU nodes, purple dots represent task-private MZI nodes, and
purple trajectories indicate normalized optical energy flow. The energy is concentrated
along structured paths rather than uniformly spread across the full mesh. The remaining
task-private MZIs are preferentially located near energy-carrying regions, supporting the
physical intuition that high-impact optical units should retain task-specific tunability while
low-damage units can be replaced by shared MPUs.
}
\label{fig:s14}
\end{figure}

\end{document}